\documentclass[twoside,twocolumn,9pt]{article}
\usepackage{extsizes}
\usepackage[super,sort&compress,comma]{natbib} 
\usepackage[version=3]{mhchem}
\usepackage[left=1.5cm, right=1.5cm, top=1.785cm, bottom=2.0cm]{geometry}
\usepackage{balance}
\usepackage{times,mathptmx}
\usepackage{sectsty}
\usepackage{graphicx} 
\usepackage{lastpage}
\usepackage[format=plain,justification=justified,singlelinecheck=false,font={stretch=1.125,small,sf},labelfont=bf,labelsep=space]{caption}
\usepackage{float}
\usepackage{fancyhdr}
\usepackage{fnpos}
\usepackage[english]{babel}
\addto{\captionsenglish}{%
  
}
\usepackage{array}
\usepackage{droidsans}
\usepackage{charter}
\usepackage[T1]{fontenc}
\usepackage[usenames,dvipsnames]{xcolor}
\usepackage{setspace}
\usepackage[compact]{titlesec}
\usepackage{hyperref}
\usepackage{epstopdf}%This line makes .eps figures into .pdf - please comment out if not required.

\definecolor{cream}{RGB}{222,217,201}

\newcommand{\lp}{\ell_\text{p}}
\newcommand{\dd}{\mathrm{d}}
\newcommand{\ee}{\mathrm{e}}

\begin{document}

\pagestyle{fancy}
\thispagestyle{plain}
\fancypagestyle{plain}{

%%%HEADER%%%
\fancyhead[C]{\includegraphics[width=18.5cm]{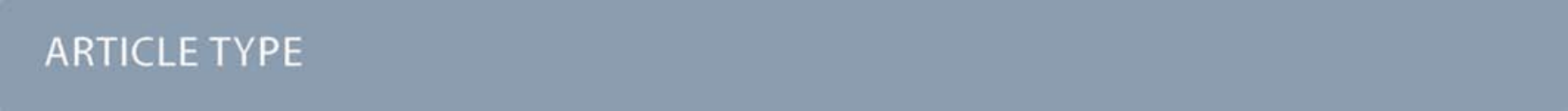}}
\fancyhead[L]{\hspace{0cm}\vspace{1.5cm}\includegraphics[height=30pt]{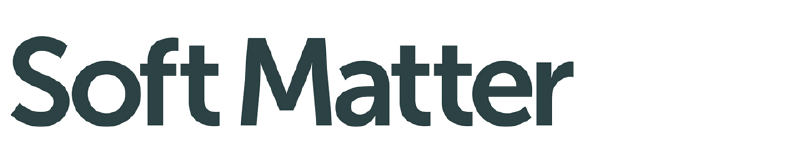}}
\fancyhead[R]{\hspace{0cm}\vspace{1.7cm}\includegraphics[height=55pt]{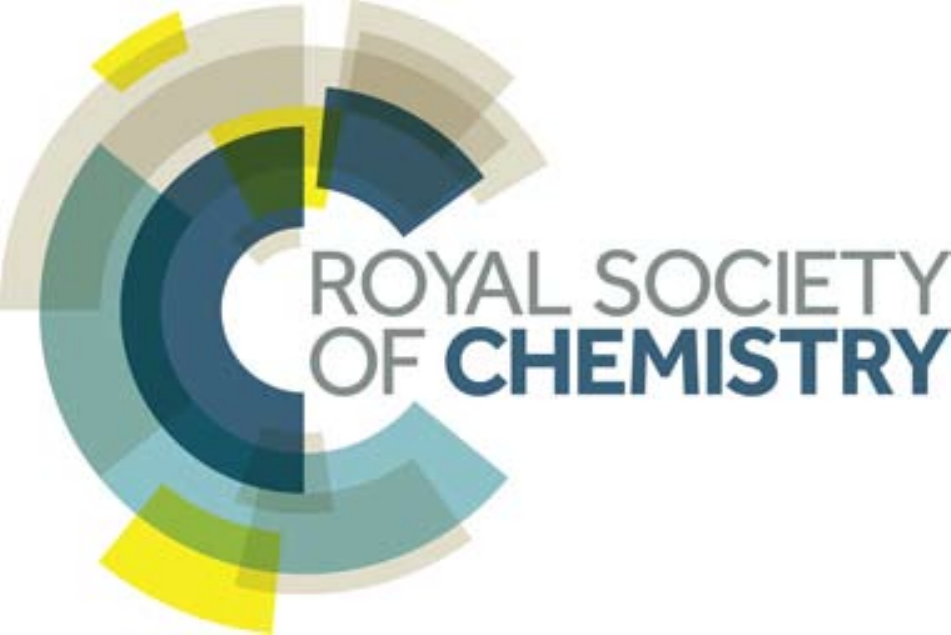}}
\renewcommand{\headrulewidth}{0pt}
}
%%%END OF HEADER%%%

%%%PAGE SETUP - Please do not change any commands within this section%%%
\makeFNbottom
\makeatletter
\renewcommand\LARGE{\@setfontsize\LARGE{15pt}{17}}
\renewcommand\Large{\@setfontsize\Large{12pt}{14}}
\renewcommand\large{\@setfontsize\large{10pt}{12}}
\renewcommand\footnotesize{\@setfontsize\footnotesize{7pt}{10}}
\makeatother

\renewcommand{\thefootnote}{\fnsymbol{footnote}}
\renewcommand\footnoterule{\vspace*{1pt}% 
\color{cream}\hrule width 3.5in height 0.4pt \color{black}\vspace*{5pt}} 
\setcounter{secnumdepth}{5}

\makeatletter 
\renewcommand\@biblabel[1]{#1}            
\renewcommand\@makefntext[1]% 
{\noindent\makebox[0pt][r]{\@thefnmark\,}#1}
\makeatother 
\renewcommand{\figurename}{\small{Fig.}~}
\sectionfont{\sffamily\Large}
\subsectionfont{\normalsize}
\subsubsectionfont{\bf}
\setstretch{1.125} %In particular, please do not alter this line.
\setlength{\skip\footins}{0.8cm}
\setlength{\footnotesep}{0.25cm}
\setlength{\jot}{10pt}
\titlespacing*{\section}{0pt}{4pt}{4pt}
\titlespacing*{\subsection}{0pt}{15pt}{1pt}
%%%END OF PAGE SETUP%%%

%%%FOOTER%%%
\fancyfoot{}
\fancyfoot[LO,RE]{\vspace{-7.1pt}\includegraphics[height=9pt]{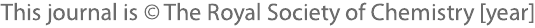}}
\fancyfoot[CO]{\vspace{-7.1pt}\hspace{13.2cm}\includegraphics{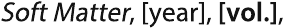}}
\fancyfoot[CE]{\vspace{-7.2pt}\hspace{-14.2cm}\includegraphics{head_foot/RF}}
\fancyfoot[RO]{\footnotesize{\sffamily{1--\pageref{LastPage} ~\textbar  \hspace{2pt}\thepage}}}
\fancyfoot[LE]{\footnotesize{\sffamily{\thepage~\textbar\hspace{3.45cm} 1--\pageref{LastPage}}}}
\fancyhead{}
\renewcommand{\headrulewidth}{0pt} 
\renewcommand{\footrulewidth}{0pt}
\setlength{\arrayrulewidth}{1pt}
\setlength{\columnsep}{6.5mm}
\setlength\bibsep{1pt}
%%%END OF FOOTER%%%

%%%FIGURE SETUP - please do not change any commands within this section%%%
\makeatletter 
\newlength{\figrulesep} 
\setlength{\figrulesep}{0.5\textfloatsep} 

\newcommand{\topfigrule}{\vspace*{-1pt}% 
\noindent{\color{cream}\rule[-\figrulesep]{\columnwidth}{1.5pt}} }

\newcommand{\botfigrule}{\vspace*{-2pt}% 
\noindent{\color{cream}\rule[\figrulesep]{\columnwidth}{1.5pt}} }

\newcommand{\dblfigrule}{\vspace*{-1pt}% 
\noindent{\color{cream}\rule[-\figrulesep]{\textwidth}{1.5pt}} }

\makeatother
%%%END OF FIGURE SETUP%%%

%%%TITLE, AUTHORS AND ABSTRACT%%%
\twocolumn[
  \begin{@twocolumnfalse}
\vspace{3cm}
\sffamily
\begin{tabular}{m{4.5cm} p{13.5cm} }

\includegraphics{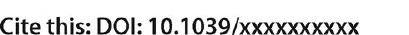} & 
\noindent\LARGE{\textbf{Curvature-dependent tension and tangential flows at the interface of motility-induced phases
}} \\%Article title goes here instead of the text
%"This is the title"...append "$^\dag$" if we want ESI
\vspace{0.3cm} & \vspace{0.3cm} \\

 & \noindent\large{Adam Patch,$^{\ast}$\textit{$^{a}$} Daniel
Sussman,\textit{$^{a}$} David Yllanes,\textit{$^{a,b,c}$} and M. Cristina
Marchetti\textit{$^{a,c}$}} \\%Author names go here instead of "Full name",etc.

\includegraphics{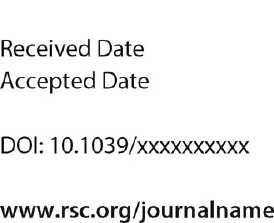} & 
\noindent\normalsize{Purely repulsive active particles spontaneously undergo
motility-induced phase separation (MIPS) into condensed and dilute phases.
Remarkably, the mechanical tension measured along the interface between these
phases is \emph{negative}.  In equilibrium this would imply an unstable
interface that wants to expand, but these out-of-equilibrium systems display
long-time stability and have intrinsically stiff boundaries. Here, we study
this phenomenon in detail using active Brownian particle simulations and a
novel frame of reference. By shifting from the global (or laboratory) frame to
a local frame that follows the dynamics of the phase boundary, we observe
correlations between the local curvature of the interface and the measured
value of the tension.  Importantly, our analysis reveals the presence of
sustained local \textit{tangential} motion of particles within a surface layer
in \textit{both} the gas and the dense regions.  The combined tangential
current in the gas and self-shearing of the surface of the dense phase suggest
a stiffening interface that redirects particles along itself to heal local
fluctuations.  These currents restore the otherwise wildly fluctuating
interface through an out-of-equilibrium Marangoni effect. We discuss the
implications of our observations on phenomenological models of interfacial
dynamics.}

\end{tabular}

 \end{@twocolumnfalse} \vspace{0.6cm}

  ] %%%opened on line 124
%%%END OF TITLE, AUTHORS AND ABSTRACT%%%

%%%FONT SETUP - please do not change any commands within this section
\renewcommand*\rmdefault{bch}\normalfont\upshape \rmfamily \section*{}
\vspace{-1cm}

%%%FOOTNOTES%%%

\footnotetext{\textit{$^{a}$~Department of Physics and Soft and Living Matter
Program, Syracuse University, Physics Building, Syracuse, New York 13244, USA}}
\footnotetext{\textit{$^{b}$~Instituto de Biocomputaci\'on y F\'isica de
Sistemas Complejos (BIFI), 50009 Zaragoza, Spain}}
\footnotetext{\textit{$^{c}$~Kavli Institute for Theoretical Physics,
University of California, Santa Barbara, CA 93106, USA}}
\footnotetext{\textit{$^{\ast}$~apatch@syr.edu}}

%Please use \dag to cite the ESI in the main text of the article.  If you
%article does not have ESI please remove the the \dag symbol from the title and
%the footnotetext below.  \footnotetext{\dag~Electronic Supplementary
%Information (ESI) available: [details of any supplementary information
%available should be included here]. See DOI: 10.1039/b000000x/} additional
%addresses can be cited as above using the lower-case letters, c, d, e... If
%all authors are from the same address, no letter is required

%\footnotetext{\ddag~Additional footnotes to the title and authors can be
%included \textit{e.g.}\ `Present address:' or `These authors contributed
%equally to this work' as above using the symbols: \ddag, \textsection, and \P.
%Please place the appropriate symbol next to the author's name and include a
%\texttt{\textbackslash footnotetext} entry in the the correct place in the
%list.}

%%%END OF FOOTNOTES%%%

%%%MAIN TEXT%%%%

\section{Introduction} 
Pierre-Gilles de Gennes wrote\cite{de2005soft} that ``the interfaces between
two forms of bulk matter are responsible for some of the most unexpected
actions...  the overlap region is \emph{mobile, diffuse, and active}.'' This
description is particularly apt as applied to the emergent behavior of dense
collections of active Brownian particles (ABPs), in which purely repulsive
particles are driven out of equilibrium via self-propulsive forces in an
overdamped environment\cite{Fily2012a}. Even in the absence of attractive
interactions, such systems can spectacularly phase separate into a dense liquid
phase coexisting with a dilute gaseous
phase~\cite{Tailleur2008,Fily2012a,Redner2013,Cates2015}. This
motility-induced phase separation (MIPS) is heuristically understood by
considering the persistent dynamics of an individual particle, and it occurs
when the time for a particle to re-orient after a collision becomes long
relative to the typical mean free time between those collisions. The occurrence
of MIPS has also been described through an approximate mapping onto an
effective equilibrium system undergoing conventional  vapor-liquid phase
separation.
\cite{Wittkowski2014,Fily2014,Solon2015,Solon2015a,Cates2015,Winkler2015,Takatori2015,marconi2016,solon2018generalized}
It has been demonstrated in numerical simulations in two and three dimensions
for various minimal models and repulsive potentials.
\cite{Fily2012a,Redner2013b,Bialke2013,Yang2014,takatori2015thermo}
Experiments in active colloids and bacterial suspension, however, generally
observe the formation of finite-size clusters rather than bulk phases,
suggesting that non-generic phenomena may be at play and arrest the phase
separation.
\cite{theurkauff2012,Buttinoni2013,Palacci2013,mallory2014,Bechinger2016,Liu2017}

\begin{figure}[t] \begin{center} 
\includegraphics[width=0.45\textwidth]{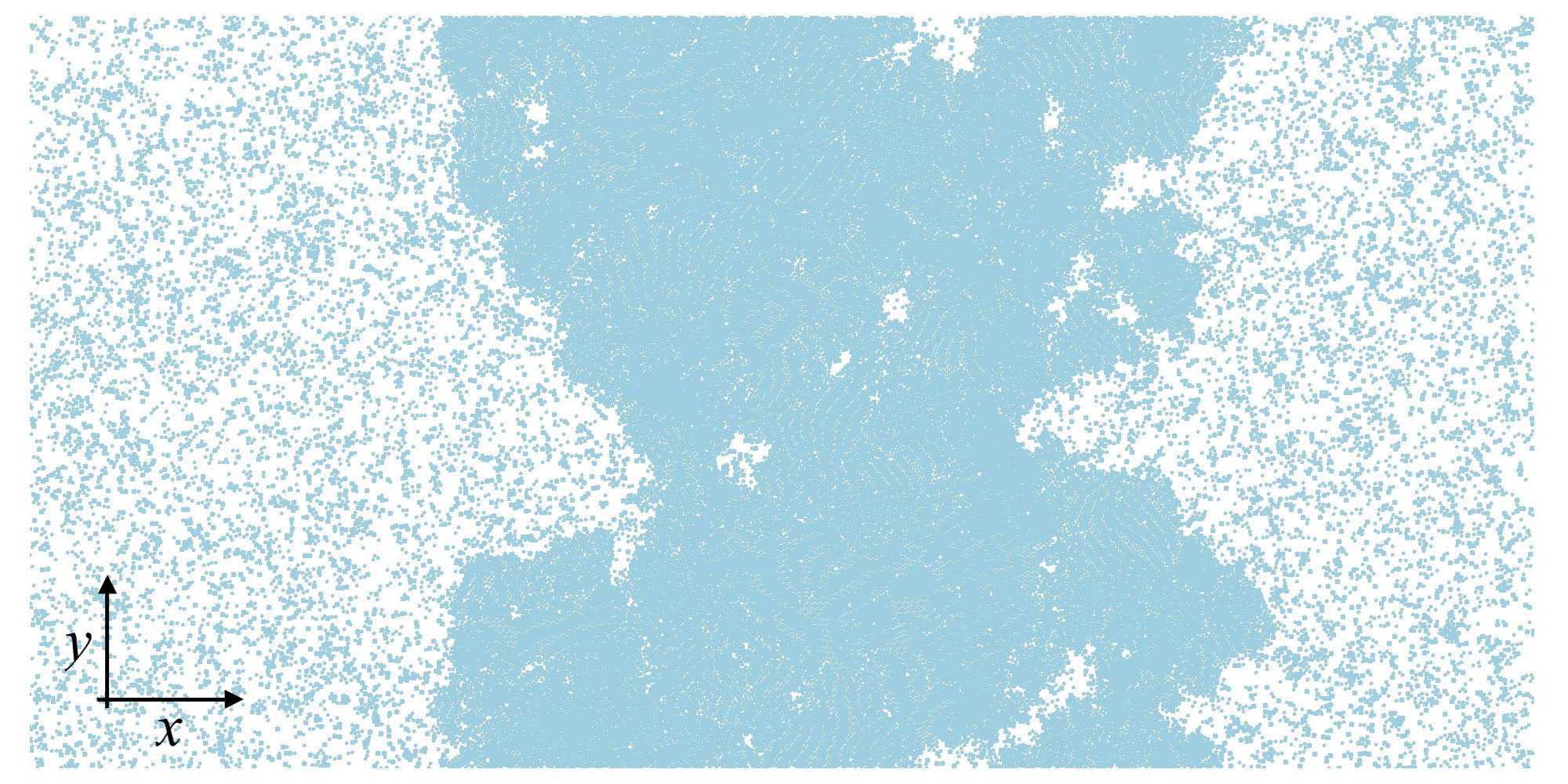}
\includegraphics[height=0.15\textwidth]{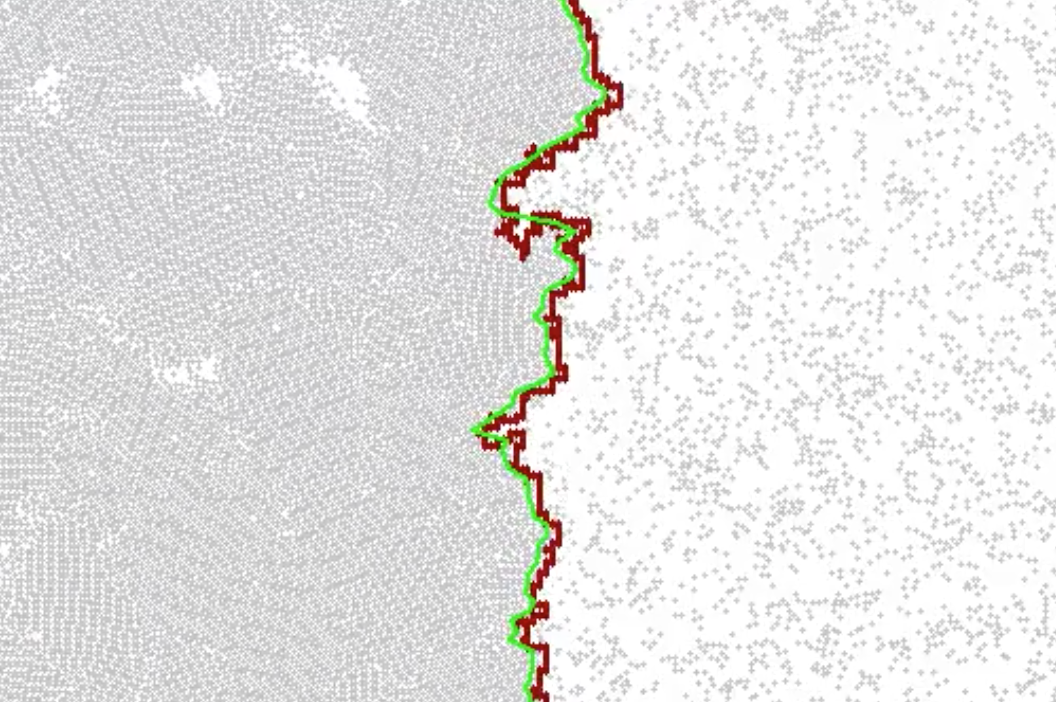}
\includegraphics[height=0.16\textwidth]{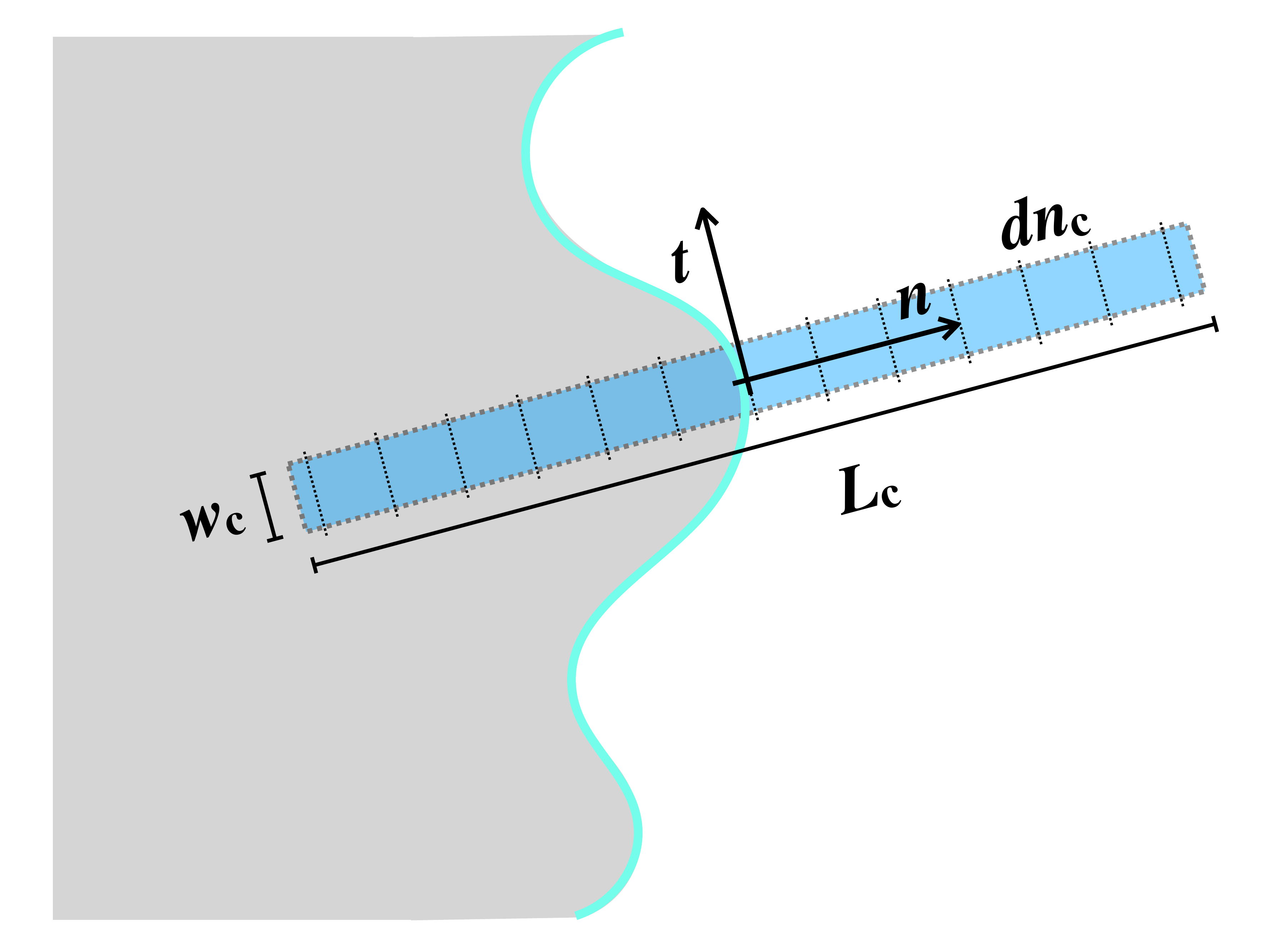}
\caption{ 
Top: A snapshot of a system composed of roughly $2\times10^5$ active Brownian
particles of radius $r_0$ undergoing spontaneous separation into dense and
dilute phases. The persistence length of the particles (defined in the text) is
$\lp = 100 r_0$; the area fraction is $\phi=0.5$ and phase separation into a
strip geometry is attained by choosing the aspect ratio of the simulation box
to be $L_x/L_y= 2$, with periodic boundary conditions (here $L_x = 1600 r_0$).
Bottom left: A demonstration of two methods for identifying the interface. The
red curve results from a contour-finding algorithm that captures all overhangs
and allows for local curvature measurements, while the green curve considers
only the outermost particles at each value of $y$ and can be used to obtain the
spectrum of the interface height fluctuations (see Appendix~\ref{App:contour}
for  details). Bottom right: A schematic of the the local frame we use to
measure dynamical quantities near the interface (see Appendix
~\ref{App:localFrame} for  details).
\label{fig:snapshot}} 
\end{center}
\end{figure}

A typical snapshot of such an out-of-equilibrium phase separation is shown in
Fig.~\ref{fig:snapshot}, where the enormous fluctuations characteristic of
MIPS are readily seen. Large fluctuations occur both at the interface and in
the bulk of the dense phase, where bubbles of the dilute phase spontaneously
nucleate and travel to the phase boundary,
breaking at the surface (see SI for a video). Despite the wildly fluctuating
nature of the interfaces,  connections with interfacial properties of
equilibrium phases have been identified.
\cite{Bialke2015,speck2016stochastic,lee2017interface,marconi2016,paliwal2017,solon2018generalized,tjhung2018reverse}
For example, as we describe in more detail below, the scaling of the
interfacial stiffness with system size is found to be consistent with
equilibrium arguments.  There is, however, a major caveat: the measured
interfacial tension $\gamma$ is
\emph{negative},\cite{Bialke2015,lee2017interface,solon2018generalized}
and the equilibrium arguments connecting it to an interfacial the stiffness
require one to take $|\gamma |$ as the relevant
quantity.\cite{speck2016stochastic}

How can we reconcile a stable, equilibrium-like interface with negative values
of surface tension, especially in a system driven by far-from-equilibrium
dynamics? In this work we rely on extensive simulations to study the structure
of the MIPS interface. We show that for such strongly fluctuating interfaces --
where the instantaneous deviation of the interfacial height from its average
value is decidedly not small -- a transformation to the local coordinate frame
along the interface (illustrated in Fig. \ref{fig:snapshot}) reveals surprising
additional structure. An important finding is a strong correlation between the
local curvature and the magnitude of the surface tension. By examining in
detail the local dynamics near the fluctuating interface, we demonstrates the
existence of surface layers with large local tangential particle motion in
\textit{both} the dilute and the dense regions. While such tangential currents
in the interfacial region within the gas have been highlighted before, the
self-shearing of the surface layer in the dense phase is a new result. This
observation suggests a new mechanism for the stabilization of active interfaces
where the interface directs particles along itself to heal fluctuations.  

Below, we first describe our simulation model and study the properties of the
interface in a global reference frame, in line with other studies.
\cite{Bialke2015,speck2016stochastic,lee2017interface,solon2018generalized}
In addition to expanding on previous results for the scaling of the interfacial
width with simulation size, we directly study the spectral density of the
interface fluctuations.  Despite the  non-equilibrium character of the system,
this spectrum is surprisingly equilibrium-like.  In this global frame we also
study the (mechanically defined) value of the interfacial tension over a wide
range of parameters, varying the persistence length of the particles. We
confirm the unusual result that the interfacial width scales with a stiffness
proportional to the absolute value of a negative $\gamma$.

We then shed new light on the mechanism of interfacial stability by shifting to
a \emph{local frame} defined along the interface, utilizing an
algorithmically traced interface contour with single particle resolution.  In
this frame, it is straightforward to define normal and transverse particle
fluxes and forces.  Using this technique we quantify the strong correlations
between the local curvature of the interface and the value of the surface
tension: although we find that the tension is always negative, it is closest to
zero in regions of large positive curvature (inward ``valleys'') and most
strongly negative in regions of large negative curvature (outward ``peaks'').

We close by presenting a heuristic picture of the emergent collective behavior
of self-propelled particles near the interface, highlighting how their dynamics
produces a negative interfacial ``tension'', resulting in an effectively
``extensile'' interface that tends to grow longer, while maintaining its
integrity. This is in sharp contrast with familiar equilibrium interfaces that
are ``contractile'' in the sense that the positive tension always tends to
shorten the interface. These interfaces have been phenomenologically
interpreted in terms of Edwards-Wilkinson-like growth
processes,\cite{edwards1982surface,lee2017interface} but the dependence of the
surface tension on local curvature naturally leads to simple KPZ-like
equations.\cite{KPZ} We attempt different scaling collapses of our data for
interfacial roughening in an effort to discriminate between these scenarios. 

\section{Model and Methods}

We simulate a system of active Brownian particles (ABPs) in a regime in which
they are known to undergo motility-induced phase separation. We choose  a strip
geometry for ease of identification of the interface. We identify the dense
phase as all members of the largest set of touching particles, while the gas
phase is composed of those remaining particles.
\subsection{Active Brownian Particle Model} A minimal model of monodisperse,
purely repulsive ABPs~\cite{Fily2012a} consists of $N$ self-propelled particles
with interaction radius $r_0$. We place these in a rectangular simulation
domain of sides $L_x$ and $L_y$ with periodic boundary conditions, setting
$L_x/L_y = 2$ so that the bulk phases yield a quasi-$1D$ interface (see
Fig.~\ref{fig:snapshot}). Each particle is identified by its position
$\mathbf{r}_i$ and director $\hat{\mathbf{e}}_i = ( \cos\theta_i ,
\sin\theta_i )$ that defines the direction of the propulsive force
$\mathbf{F}_i^\text{s} = (v_0/\mu) \hat{\mathbf{e}}_i$, where $v_0$ is the
propulsion speed and $\mu$  the mobility (inverse friction). The particles are
governed by the overdamped Langevin equations
\begin{align} 
\dot{\mathbf{r}}_i &= \mu \big( \mathbf{F}_i^\text{s} +
\sum_{j\neq i} \mathbf{F}_{ij} \big )\;, \\ \dot\theta_i &=
\sqrt{\frac{2}{\tau_\text{r}}} \eta_i(t)\;, 
\end{align} 
where $\eta_i(t)$ is a Gaussian random torque  with zero mean and variance
$\langle\eta_i (t)\eta_j (t')\rangle = \delta_{ij}\delta(t-t')$. The variance
of this fluctuating torque is set by the persistence time $\tau_\text{r} =
1/D_\text{r}$, where $D_\text{r}$ is rotational diffusion. The pair forces
between particles $i$ and $j$, $\mathbf{F}_{ij}$, are obtained from a repulsive
Weeks-Chandler-Anderson potential, 
$\mathbf{F}_{ij} = - \frac{\partial
V_\text{WCA}}{\partial r_{ij}} = \frac{12
\epsilon}{r_\text{c}} \big [ \big ( \frac{r_\text{c}}{r_{ij}} \big )^{13} 
						- \big (
						  \frac{r_\text{c}}{r_{ij}}
\big )^{7} \big ] \hat{\mathbf{r}}_{ij}$ for $r < r_\text{c} = 2 r_0$ and
$|\mathbf{F}_{ij} | = 0$ otherwise, where $\mathbf{r}_{ij}=\mathbf{r}_i-\mathbf{r}_j$ is
the interparticle separation and
$\hat{\mathbf{r}}_{ij}=\mathbf{r}_{ij}/|\mathbf{r}_{ij}|$. 

We neglect noise in the translational dynamics, which is less important than
the orientational noise in both synthetic active colloids and swimming
bacteria.~\cite{marchetti2016minimal,Liu2017}  The non-equilibrium nature of
this active model is provided entirely by the propulsive force
$\mathbf{F}^\text{s}_i$ of each particle. After integrating out the angular
dynamics, $\mathbf{F}_{i}^\text{s}$ represents a non-Markovian stochastic force
correlated over the persistence time $\tau_\text{r}$. Since the finite
correlation time of the noisy propulsive force is not matched by similar
correlations in the (constant) mobility $\mu$, the system does
not obey the fluctuation-dissipation theorem embodied by the
Stokes-Einstein relation.  

The persistence time $\tau_\text{r}$ controls the crossover from ballistic to
diffusive single particle dynamics: at short times the dynamics of
non-interacting ABPs is ballistic, and for $t\gg\tau_r$ it is diffusive with
diffusion coefficient $D_\text{s}=v_0^2\tau_r/2$. The single-particle dynamics
can be characterized by the persistence length $\lp = v_0 \tau_\text{r}$,
which, together with the area fraction $\phi$, controls the phase behavior of
this interacting non-equilibrium system.~\cite{Redner2013,marchetti2016minimal} 

\subsection{Capillary waves and interface width}
Taking a mesoscopic view of the interface, one can characterize the
fluctuations in terms of the deviations of the instantaneous location of the
surface along the $x$ direction from its mean value,~\cite{rowlinson2013molecular}
$\delta h(y,t) =h(y,t)-\bar{h}(t)$, with $\bar{h}(t)=\frac{1}{L_y}\int_0^{L_y}
dy \; h(y,t)$.  The mean interfacial width can then be written as, 
\begin{equation}
  w^2 = \frac{1}{L_y} \int_0^{L_y} \dd y \; \langle | \delta h(y) |^2 \rangle 
 = \sum_q \langle |\delta h(q) |^2 \rangle\;,
\label{eq:w2_1}
\end{equation}
where
\begin{align}
  \delta h(q) &= \frac{1}{L_y} \int_0^{L_y} \dd y \;\delta h(y) \ee^{-\text{i}q y}.
\end{align} 
In thermal equilibrium, interfaces carry an excess free energy $E_\text{s} =
\gamma \ell$ determined by the constant interfacial tension, $\gamma$,
and the length $\ell$ of the interface, with
\begin{equation}
  \ell = \int_0^{L_y} \dd y \; \sqrt{1+|\nabla_y h(y)|^2} \approx L_y \left [ 1 + \frac{1}{2} \sum_q q^2 |\delta h(q)|^2 \right] \;.
\end{equation}  
The interfacial height mode amplitudes are then determined by the equipartition theorem as
\begin{equation}
  \langle | \delta h(q)|^2 \rangle = \frac{2}{L_y}\frac{kT}{\gamma q^2}\;.
\end{equation}
Eq.~(\ref{eq:w2_1}), we can immediately calculate the interface width as
\begin{equation}
  w^2 = w_0^2 + \sum_{q>0}\langle |\delta h(q)|^2\rangle = w_0^2 + \frac{L_y}{12 \sigma} \;,
\label{eq:w2L}
\end{equation} 
where $w_0^2$ describes the fluctuations of the $q=0$ mode
and $ \sigma=\gamma/kT$ is the interfacial stiffness that measures the cost of
deformations along the entire length $L_y$.

\subsection{Interfacial tension}

A mechanical definition of the interfacial tension $\gamma$ can be obtained by
examining the work  $\delta W = \gamma \delta \ell$ needed to change the length
of the interface by an amount  $\delta \ell$.  Here we follow the standard
quasi-thermodynamic treatment of Ref.\cite{rowlinson2013molecular}, where in
equilibrium it is shown that this mechanical definition yields the same value
as that obtained from  the interfacial fluctuations. As we will see, this is
not, however, the case for ABPs~\cite{Bialke2015, lee2017interface}.

Working in two dimensions, we consider a one-component system confined to a box
of area $A=\ell^2$ and separated into two bulk phases, with a vertical
interface at some position $0<x_0<\ell$. If the area of the system is changed
isotropically by an amount $\delta A$, the associated work is controlled by the
pressure, with $\delta W = - p \delta A$.  To define the tension we imagine
isothermally and reversibly deforming the sides of the confining box so that
the interface increases in length by an amount $\delta \ell$, while maintaining
fixed area. This requires an anisotropic deformation of the box, but the
symmetry of the interface ensures that the pressure tensor can only have
non-zero components $p_{xx}$ and $p_{yy}$.  The tangential work
done to the system in increasing the length of the interface is then
\begin{equation}
   \delta W_t = -\delta\ell \int_0^\ell \dd x p_{yy}(x) 
\end{equation}
and the normal work done in keeping the area fixed is
\begin{equation}
   \delta W_n =\ell \delta \ell p_{xx}. 
\end{equation} 
In writing the above expression we have assumed mechanical stability of the
interface, $\nabla\cdot \mathbf{p} = 0$. This, together with the symmetry
of the system ensures that $p_{xx}$ is not itself a function of $x$ or $y$. The
total work done is then 
\begin{equation}
   \delta W =  \delta W_n+ \delta W_t =\delta\ell \int_0^\ell \dd x \left(  p_{xx} - p_{yy}(x) \right)\;.
\end{equation} 
Comparing this expression with $\delta W = \gamma \delta \ell$ we recover
the Kirkwook-Buff expression\cite{kirkwood1949statistical},
\begin{equation}\label{eq:tension} 
  \gamma = \int_\text{dense}^\text{dilute} \dd x\ \big [ p_\text{n} -
p_\text{t}(x) \big ],  
\end{equation} 
where we have more generally replaced Cartesian components of the tensor with
normal and tangential components, and assumed that the two phases we are
considering are a dense and dilute phase.

Equation \ref{eq:tension} quantifies the surface tension as the total
anisotropy in pressure across the interface. It assumes of course that the
local pressure tensor has a mechanical definition. In the next section we
outline how we measure the local pressure tensor for our active system and
compute Eq. \ref{eq:tension}.

\subsection{Pressure in Active Matter}
The pressure $p$ of a system of overdamped torque-free active particles
involves two contributions: a contribution due to direct particle interactions,
$p_\text{d}$, and a ``swim'' contribution that represents the flux of
propulsive forces across a unit area, $p_\text{s}$, with $p = p_\text{d}+
p_\text{s}$. The components of the interaction pressure tensor
$p_\text{d}^{\alpha\beta}$, where $\alpha,\ \beta$ denote Cartesian indices,
are given by a virial expression in terms of  particle pair interactions,
 \begin{equation}
p_\text{d}^{\alpha\beta} = \frac{1}{dA} \sum_{i,j} {F}_{ij}^\alpha
{r}_{ij}^\beta\;.
 \end{equation} 
The swim pressure can also be calculated via a virial-like
expression.~\cite{Yang2014,marchetti2016minimal,patch2017kinetics} A
numerically more useful expression for the local the swim pressure is given in
terms of the flux of ``active impulse'', as
 \begin{equation} 
 p_\text{s}^{\alpha\beta} =
\frac{1}{dA} \sum_{i} {J}_i^\alpha {v}_{i}^\beta\;,
\label{eq:ps} 
\end{equation}
\noindent
where $\mathbf{v}_i$ is the velocity of particle $i$ and
$\mathbf{J}_i[\theta_i(t)] = \mathbf{F}^s_i(t) \tau_\text{r}$ is the active
impulse introduced in Ref. \cite{fily2017mechanical}, where it was shown that
the expression given in Eq.~\eqref{eq:ps} for the swim pressure is equivalent
to the virial one proposed in previous work.\cite{Yang2014,Takatori2014} The
form in term of active impulse is more convenient for numerical studies
because it avoids the strong finite size effects that arise in calculations of
the virial expression for the swim pressure.\cite{patch2017kinetics}
\begin{figure} [t]
\includegraphics[height=0.45\textwidth,angle=270]{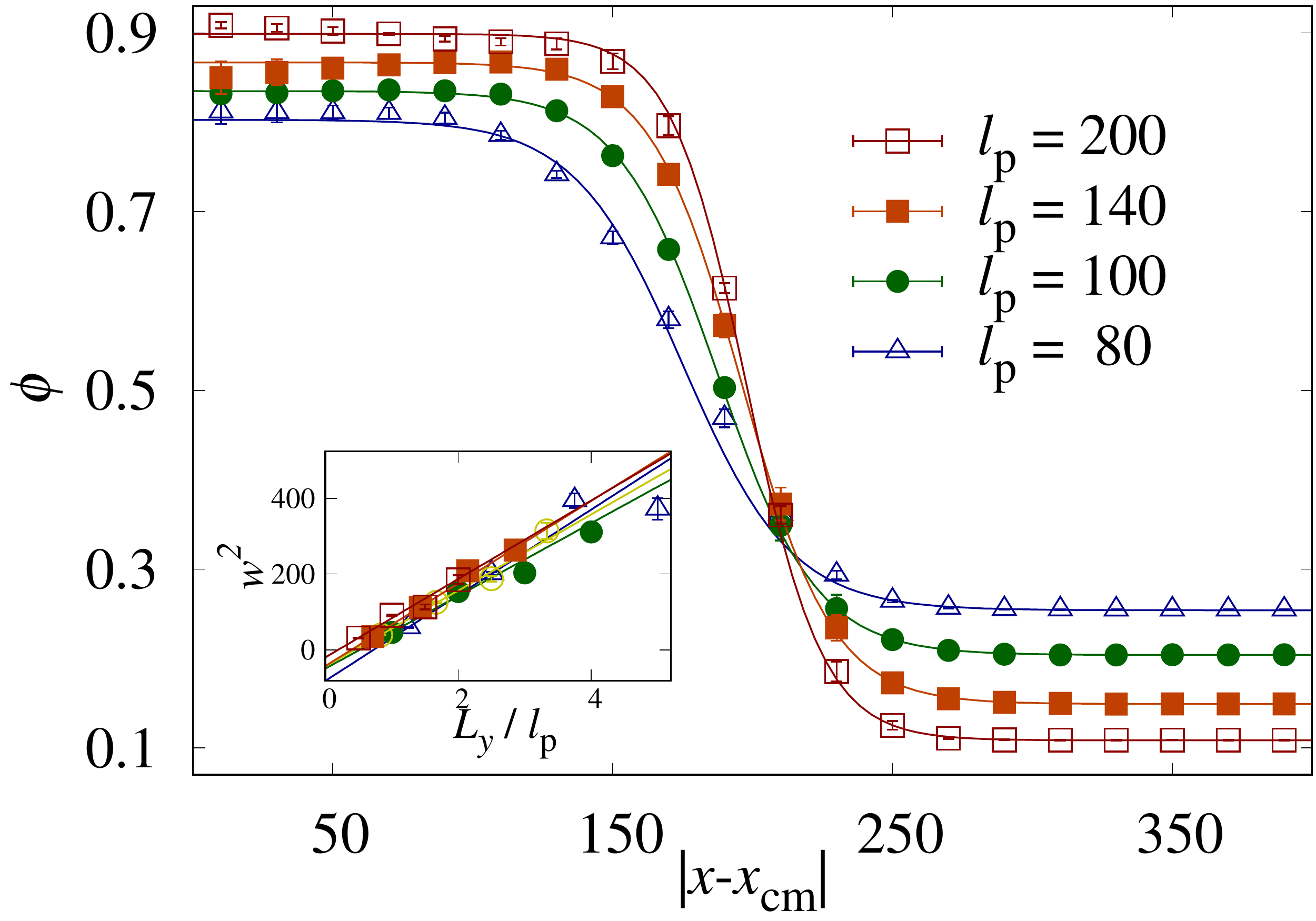}
\includegraphics[height=0.45\textwidth,angle=270]{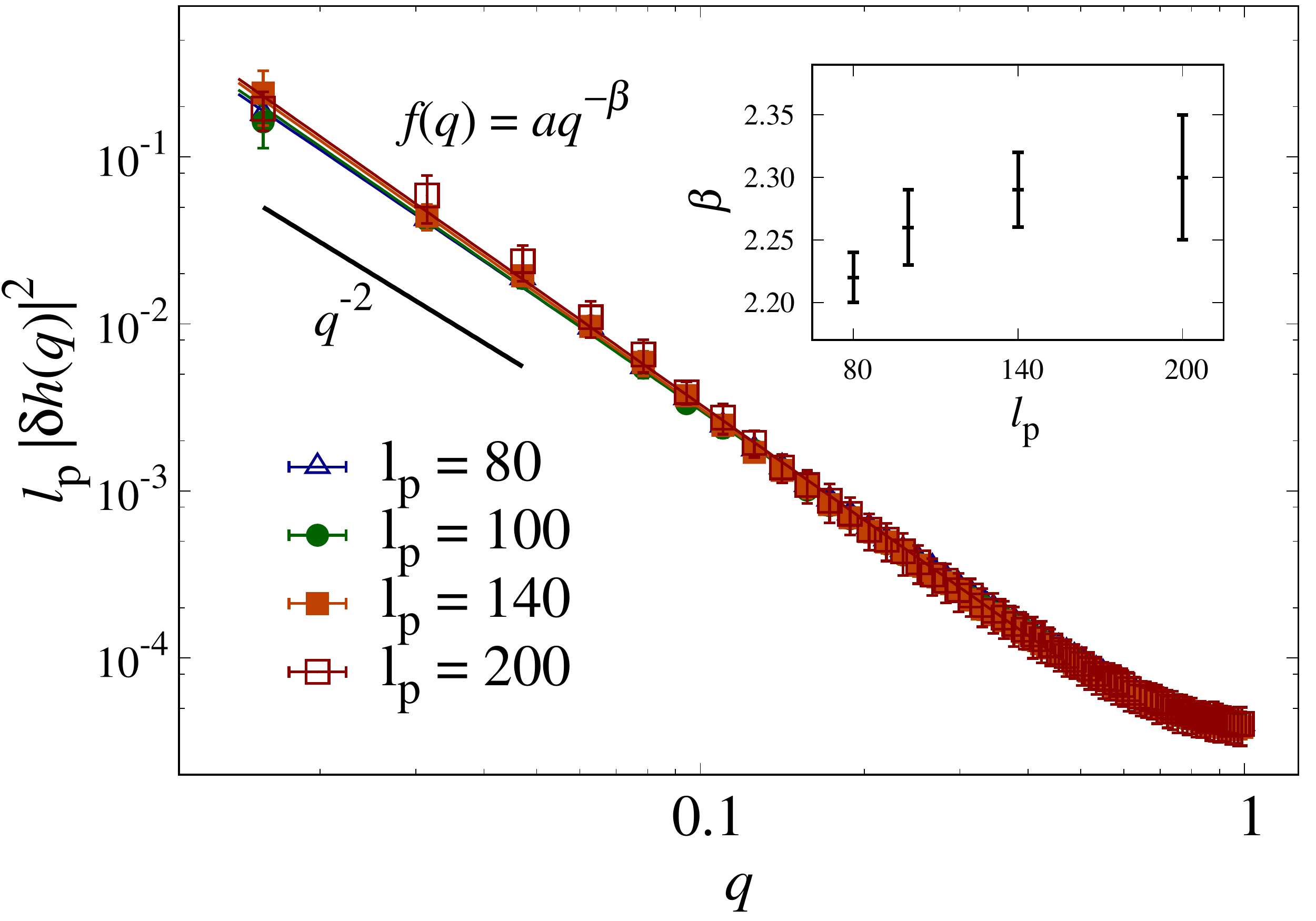} \caption{
Top: density profiles in the interface frame, for varying persistence.  [Inset]
Scaling of $w^2$ as a function of $L_y/\lp$, for comparison to
Eq.~\ref{eq:w2L}.  Bottom: Interface height fluctuation spectrum for systems
with $L_y=400$, collapse for several values of $\lp$. The equilibrium
expectation $\langle |\delta h(q)|^2\rangle \sim q^{-2}$ is shown as a
solid black line. The collapse of $\lp |\delta h(q)|^2$, signifies the clear
dependence of interfacial fluctuations on persistence. [Inset] Best fit
exponent characterizing the initial decay of the interface modes as a function
of $\lp$. 
\label{fig:phi_dh}} 
\end{figure}

\subsection{Simulations} 
Simulations of large phase-separating systems of ABPs require total running
times $t\gg\tau_\text{r}$ and integration timesteps  $\Delta
t\ll\tau_\text{r}$. \cite{patch2017kinetics} This, in addition to the
wildly fluctuating nature of the interfaces and the need to sweep different
$\ell_\text{p}$ values, means that considerable computational effort must be
made to gather an adequate statistical sample. Most of our simulations have
been carried out on GPUs using the HOOMD-blue\cite{hoomd-blue,hoomd-blue0}
simulation package, complementing with CPUs for the smaller system sizes.

To suppress transients,  we nucleate the phase-separated strip by initially
placing particles in a triangular lattice with interparticle distance $r_{ij} =
2 r_0$.  We performed  parameters sweep by using persistence lengths $\ell_p =
\{ 60, 80, 100, 120, 140, 200 \}$ and system sizes $L_x = \{200, 400, 600, 800,
1600\}$.  The area fraction was fixed at $\phi=0.5$, with  $\mu=1$ and
$v_0=100$.  We have simulated for $t \geq 1\times10^3\tau_\text{r}$, with  a
number of independent runs $N_\text{runs}=100$  for our smallest systems, down
to $N_\text{runs}=8$ for $L_x=800$.  Unless stated otherwise, the results in
this paper  refer to $L_x=800$ systems. In order to study the interface growth,
we have also run shorter simulations ($t=100\tau_\text{r}$) that do not reach
the stead state, for system sizes ranging from $L_x=400$ ($N_\text{runs}=400$)
to $L_x=2400$ ($N_\text{runs}=28$).

We note that previous studies have found strong finite-size effects when
studying the pressure of phase-separated systems with $\lp\sim L$, even at low
density.\cite{patch2017kinetics}.  Measuring the swim pressure with the active
impulse flux, Eq.~(\ref{eq:ps}), rather than  with a virial expression,
mitigates this effect.  Nevertheless, we have kept the values of $\ell_p$
smaller than $L_x$ in our work.

\begin{figure}[t] 
\includegraphics[height=0.44\textwidth,angle=270]{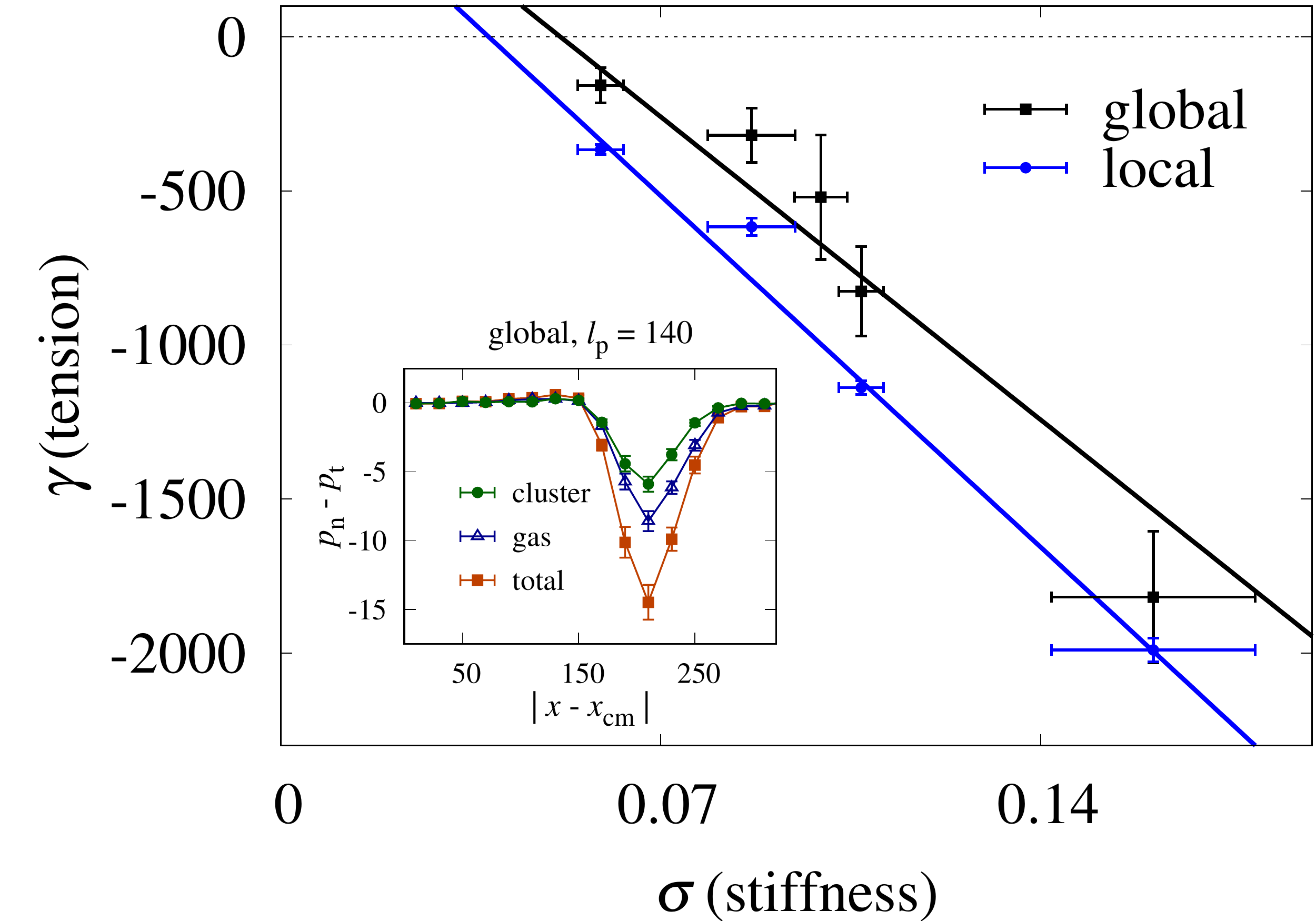}
\caption{
The main figure correlates the interfacial tension and the stiffness,
showing that the tension becomes more negative with increasing stiffness. This
suggests that the same dynamics that produces a positive stiffness yields a
negative interfacial tension. The black points correspond to measurements in
the global frame (described in Sec.~\ref{sec:globalframe}), while the blue
points correspond to measurements in the local frame of the interface
(described in Sec.~\ref{sec:localframe}).  The inset shows the difference
between normal and tangential pressure that determines the integrand in
Eq.~\ref{eq:tension}, broken down into cluster and gas contributions,
highlighting the significant net-negative contributions from both cluster and
gas particles.  
\label{fig:pNpT}} 
\end{figure}

\section{Measurements in the global frame} 
\label{sec:globalframe}
The MIPS interface is characterized by wild fluctuations (as seen in Fig.
\ref{fig:snapshot}). Upon  time-averaging, however, the interface appears
deceptively well behaved and equilibrium-like. In this section we show this by
characterizing the average density profile, determining the stiffness of the
interface from the scaling of the interfacial width, and investigating the
spectrum of interface fluctuations. We will often refer to the system as
divided into two sets of particles: those belonging to the largest connected
cluster (dense) and those in the gas (dilute).

\subsection{Interface width and stiffness}

Working in the global frame, we first consider the average area fraction projected
onto the $x$-axis, $\phi(x) = \frac{1}{L_y} \int_0^{L_y} \dd y\ \phi(x,y)$. The
resulting profile is well described by
 \begin{equation} \phi(x) = \frac{\phi_+
+ \phi_-}{2} + \frac{\phi_+ - \phi_-}{2}\tanh\bigg(\frac{x-x_0}{2w}\bigg)\;,
\end{equation}
where $\phi_\pm$ denotes the area fraction for the dense and dilute phases,
$x_0$ is the mean position of the interface, and $w$ is a measure of the
interfacial width.  In broad agreement with previous
observations,\cite{Cates2015} we find that with increasing $\lp$ the
interfacial width decreases and the difference in density between the
coexisting phases increases, as we show in the upper panel of Fig.
\ref{fig:phi_dh}.  The inset to the upper panel of Fig. \ref{fig:phi_dh} also
shows that  $w^2\sim L_y$ for a variety of $\lp$.

We note that the finite intercept at $L_y=0$ provides an estimate
of the minimum system size needed to observe MIPS in these systems. 
As shown in Section 2, Eq.\ref{eq:w2L} connects the scaling of the interfacial
width with system size $L_y$ to the intrinsic stiffness of the interface,
$\sigma$ that can be extracted from the slope of the linear curves in the inset.

\subsection{Interface fluctuations}

Although the MIPS interface is extremely rough and is characterized by frequent
overhangs, to quantify  the structure of the interfacial fluctuations we
approximate the interface by constructing a height map $h(y)$ as described in
Appendix~\ref{App:contour}. We then examine the Fourier spectrum of $\delta
h(y) = h(y) - \bar{h}$, where $\bar{h}$ is the instantaneous value of the mean
position of the height map, and average the resulting mode spectrum $\langle
|\delta h(q)|^2 \rangle$ over time in the steady state. The spectrum of
interfacial fluctuations  shown in Fig.~\ref{fig:phi_dh} is well described by
$\langle |\delta h(q)|^2 \rangle \sim \lp^{-1}q^{-\beta}$, in keeping with our
finding that the interfacial width itself scales with $\lp$. We do observe
small deviations from the strictly equilibrium expected scaling of $\beta = 2$
(see inset of lower panel of Fig.~~\ref{fig:phi_dh}).

Interfacial fluctuations have recently been measured in a mixture of active and
passive disks with attractive interactions.\cite{Vaikuntanathan2018} In that
case the attractions stabilizes the interface and activity enhances the
stiffness that grows linearly with $\lp$.

\subsection{Mechanical surface tension} 
As discussed in Section 2, a straightforward mechanical measurement of the
surface tension involves integrating differences in the local pressure tensor
across the interface,\cite{Bialke2015} via $\gamma = \int \dd x \big (
p_\text{n} - p_\text{t} \big )$ (Eq.~\ref{eq:tension}). In the global frame one
simply has $p_\text{n} = p_{xx}$ and $p_\text{t} =p_{yy}$. At an equilibrium
liquid-gas interface the positive surface tension arises from the
\textit{lowering} of tangential pressure associated with the weaker binding of 
liquid surface molecules as compared to bulk liquid molecules. In our active
system, in contrast, we find a large \textit{increase} of tangential pressure
at the interface, as shown  in the inset of Fig.~\ref{fig:pNpT} (see also
Fig.~\ref{fig:breakdown} for further details).  This large tangential pressure
is responsible for the negative value of the interfacial tension. It arises not
only from the swim pressure of the gas (Fig.~\ref{fig:breakdown}), as shown in
previous work,\cite{Bialke2015,lee2017interface,solon2018generalized} but also
from continuous tangential self-shearing motions of particles at the surface of
the dense phase.  In other words, particles moving tangentially in the
interfacial region \textit{both inside and outside} the dense phase contribute
the the negative sign of the tension, as explicitly demonstrated below.

Interestingly, we also find that the \emph{magnitude} of the surface tension
increases with persistence length, meaning that in these systems stiffer,
sharper interfaces correspond to more negative values of tension. Classically,
a negative interfacial tension would indicate an interface that prefers to
grow; here the opposite occurs. Refs.~\cite{Bialke2015,speck2016stochastic}
conjectured a ``housekeeping work'' $w_{\text{hk}}$ that accounts for the work
done \emph{by} the particle as opposed to the work done \emph{to} the particle
$w_\text{ex}$, resulting in a  relationship between tension and stiffness of
the form $\gamma = -\sigma F^\text{s} \lp$. Earlier we found that $\sigma$
is linear in $\ell_\text{p}$, implying a quadratic dependence of
$\gamma$ on $\ell_\text{p}$ (or on $\sigma$).  However, the fact that
$\sigma(\ell_\text{p})$ has a non-zero axis intercept (Fig.~\ref{fig:phi_dh})
makes it very difficult to verify this relation quantitatively.  Indeed, while
our results  show qualitative agreement with the idea that $F^\text{s} \lp$
sets an energy scale relevant to determining interfacial stability, with our
error bars $\gamma$ seems to be adequately represented by a linear dependence
on $\sigma$ (Fig.~\ref{fig:pNpT}).

\section{Measurements in the local frame} 
\label{sec:localframe}
It is well known that even in an ideal gas of ABPs the geometry of any
confining wall has a strong influence on both the structure and dynamics of
the system.\cite{Fily2014a,fily2017equilibrium} In MIPS the gas particles
interact with an emergent, self-generated boundary that continuously  absorbs
and releases particles, with zero net flux at steady state, but substantial
local tangential currents.  It is therefore illuminating to probe the
interfacial structure and dynamics with respect to a local frame at each point
of the interface. As described in more detail in Appendix~\ref{App:contour}, we
use a contour-finding algorithm to move beyond the height-map representation
$h(y)$ to the position of the interface with respect to the contour length
along the interface itself, $h(s)$. To do so, we define a local frame whose
origin is set at points along the contour and whose orientation is set by the
local normal $\mathbf{n}$ (which is also calculated using this contour by
fitting a region around a given point to a quadratic function and evaluating
its curvature). We then compute the average value of the pressure tensor in
slices of fixed width at each point along the interface.  Although our results
are quantitatively sensitive to the choice of slice width and curvature
coarse-graining scale, we have confirmed that our qualitative results are
robust to any sensible choice of these parameters.

\subsection{Dynamics along the interface}

We first illustrate our findings by looking at the structure and dynamics of
the particles in this local frame of measurement. The stability of MIPS
interfaces has often been heuristically explained by particles pointing
``inward'' at the outer edge of the dense cluster, with a rotational
re-orientation time for these particles that is longer than the typical time
for a gas particle to arrive at the interface.\cite{Cates2015} The condition of
net zero flux then suggests large transverse currents in the gas phase outside
the dense cluster.\cite{Vaikuntanathan2018}

\begin{figure}[t] 
\includegraphics[height=0.48\textwidth,angle=270]{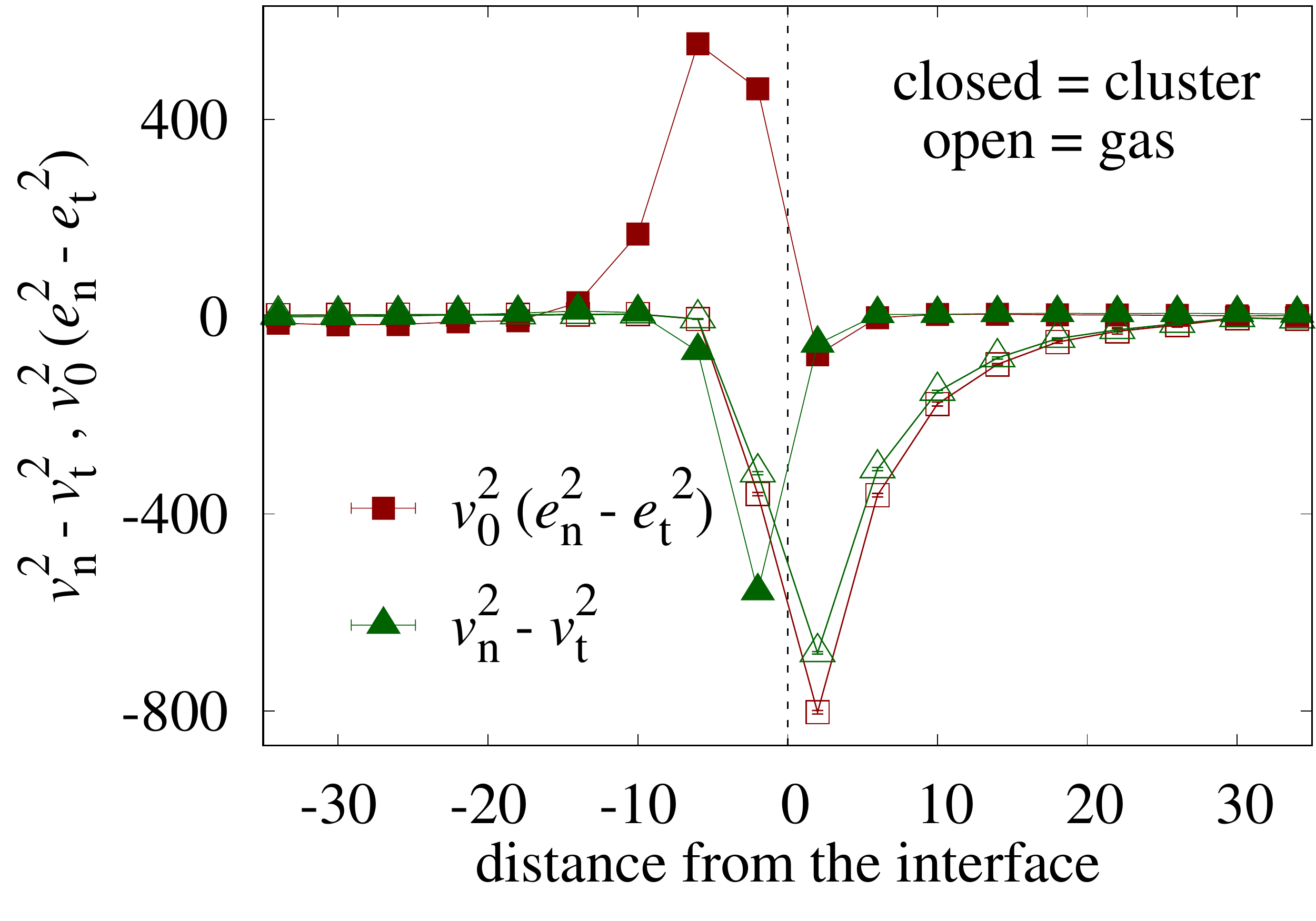} 
\caption{
We use the local frame along the phase boundary to illuminate the complex
dynamics near the interface---in particular the anisotropy of director and
velocity. The $x$-axis shows the distance from the interface
measured along the normal, as shown in Fig.~\ref{fig:snapshot}.  
Near the
interface, the average director field shows an excess of inward pointing swim
force just inside the cluster and an excess of transverse swim force just
inside the gas. The anisotropy of the resulting motion shows that the gas moves
with its director field, but the cluster self-shears as its surface particles
move tangentially.
\label{fig:local_dynamics}
}
\end{figure}

In Fig. \ref{fig:local_dynamics} we can see this dynamics at play in the local
frame of the interface. The square points show the average anisotropy of the
director, $e_\text{n}^2-e_\text{t}^2$, and show that indeed particles in the
dense cluster but near the interface preferentially point inwards, while
particles in the gas phase preferentially point tangent to the local interface.
An examination of the velocity field itself, $v_\text{n}^2-v_\text{t}^2$,
reveals, however, an unexpected behavior. In the gas phase the velocity field
and the director point in the same direction as interactions are negligible,
but they are distinctly different for particles at the interface within the
dense phase. One clearly sees that there is a local transverse current even in
the dense phase, as the projection of the swim force in the tangential
direction causes the clustered particles to slide along the very boundary they
are defining.

The transverse swim current is also enhanced by persistence, which we have
shown corresponds to stiffer interfaces. Additional details are given in
Appendix~\ref{App:local_dynamics} where we examine the behavior in the
interfacial layers both within the dense and the gas phases.  One observes
tangential particle motions in both regions and an associated local stiffening
of the interface, with the fluctuating boundary providing a local guiding
effect on active gas particles similar to that observed for curves solid walls.
\cite{fily2017equilibrium}. 

\subsection{Interfacial curvature and local surface tension}
The findings above, showing correlated flows in both the dense and dilute
phases on either side of the interface, qualitatively suggest a Marangoni-like
local mass transport that stabilizes the interface. Inspired by this, we
explicitly investigate the connection between the curvature of $h(s)$ and the
\emph{local}  mechanical measurement of surface tension. 

\begin{figure}[t] 
\includegraphics[width=0.45\textwidth]{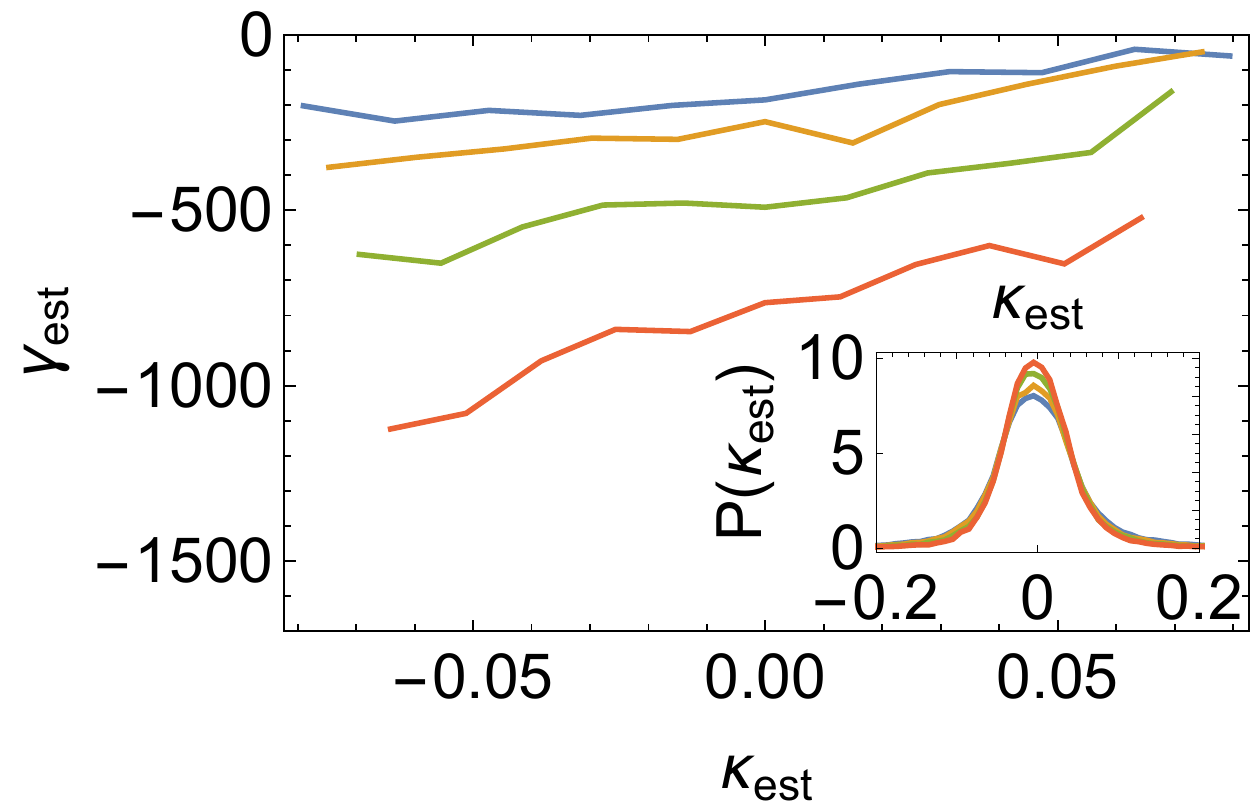} 
\caption{Correlation between the estimated local surface tension and local
curvature for $\lp = 80,\ 100,\ 140,\ 200$ (top to bottom). Negative and
positive curvature values correspond to mountains and valleys of the interface,
respectively. The correlation between curvature and tension provides a
Marangoni-like effect which allows the interface to stabilize itself. [Inset]
Probability distribution of local curvatures for the same parameters (where
higher values of $\lp$ have a more peaked distribution of curvature).
\label{fig:curvature}} 
\end{figure}

In the local frame, we still find on average a negative value for the surface
tension. In this frame, though, we find that the spatial profiles of
$p_\text{n} - p_\text{t}$  are much sharper, and individual slices can be
binned according to the local value of curvature.  These measurements have
enormous fluctuations (with a variance at least an order of magnitude larger
than the mean), but careful averaging allows us to distinguish a clear
correlation between local curvature and local tension, which we show in Fig.
\ref{fig:curvature} (restricting ourselves to values of the local curvature
within one standard deviation of the mean to ensure sufficient statistics).
Although the mean surface tension is negative for all values of curvature, we
find that the outward bulges of the interface (i.e., regions of negative
curvature) are quantitatively more unstable than the inward valleys. This
gradient of stability provides a mechanism for the interface to dynamically
stabilize itself as transverse currents from the bulges on average fill in the
gaps in the valleys.

\section{Discussion}

In a model of purely repulsive active Brownian particles undergoing motility
induced phase separation, we have explored the surprising dichotomy of an
interface that on one hand exhibits some equilibrium-like properties (a
well-behaved time-averaged density profile and a nearly $q^{-2}$ spectrum of
interfacial fluctuations), but on the other is governed by strong fluctuations
driven by non-equilibrium physics, resulting in negative interfacial tension.
It is tempting to try to write down a phenomenological model of the interface.
Some authors have in fact proposed mapping the MIPS interface to an
Edwards-Wilkinson (EW) growth model.\cite{lee2017interface} On the other hand,
our finding that the surface tension is itself a function of the local
curvature  naturally leads to additional terms, such as those in the
Kardar-Parisi-Zhang (KPZ) description.\cite{KPZ}

Although it is clear that in the presence of a negative surface tension
additional terms would be needed to stabilized the KPZ equation, we have tried
to discriminate between different universality classes for the MIPS interface
by measuring the critical exponents characterizing interfacial growth and
steady state fluctuations. For $1d$ interfaces both the EW and KPZ models are
characterized by the same roughening exponent, $\alpha=1/2$, characterizing the
growth of the steady-state interface width with system size $L_y$. They differ,
however, in the exponent $\beta$ that controls  the interface growth at short
times, $w(t)\propto t^\beta$, with $\beta_\text{KPZ}=1/3$ and
$\beta_\text{EW}=1/4$. Discriminating between the two scenarios is numerically
challenging for two reasons. First, the values of $\beta$ differ only slightly
in the two models. Second, as previously pointed out, there is a
\emph{smallest} system size necessary to observe MIPS, so achieving even a
single decade in linear system size $L_y$ requires very large computational
effort. Nevertheless, we have run many short simulations, starting from a flat
interface, to try to characterize the growth process. Surprisingly, we find
that the EW dynamical exponent seem to better collapse the data than the (from
a symmetry standpoint) more natural KPZ model.  In particular, fits to $w(t) =
A t^\beta$ for our largest systems suggest $\beta=0.23(3)$.

\begin{figure}[h] 
\includegraphics[width=0.48\textwidth]{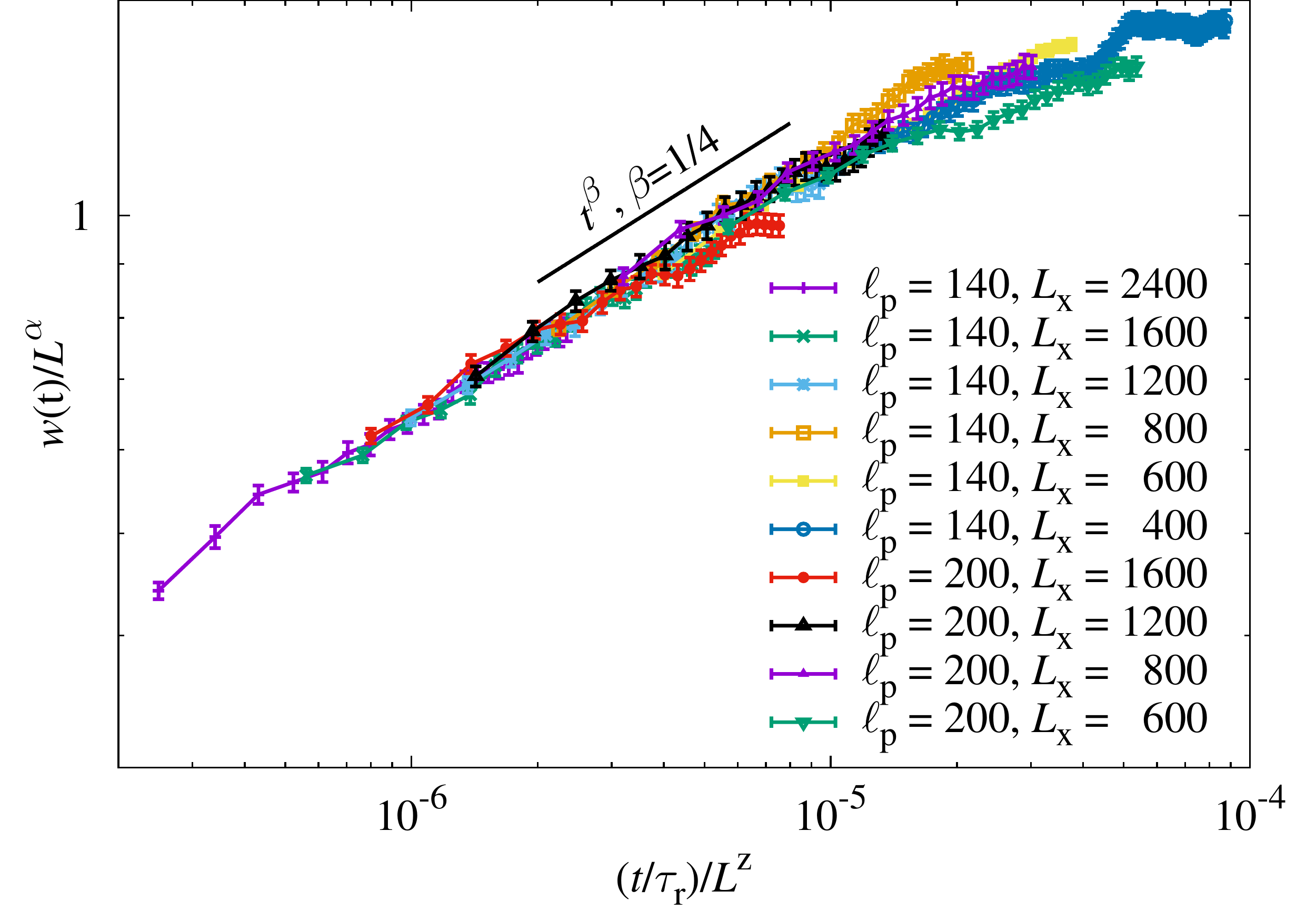} 
\includegraphics[width=0.48\textwidth]{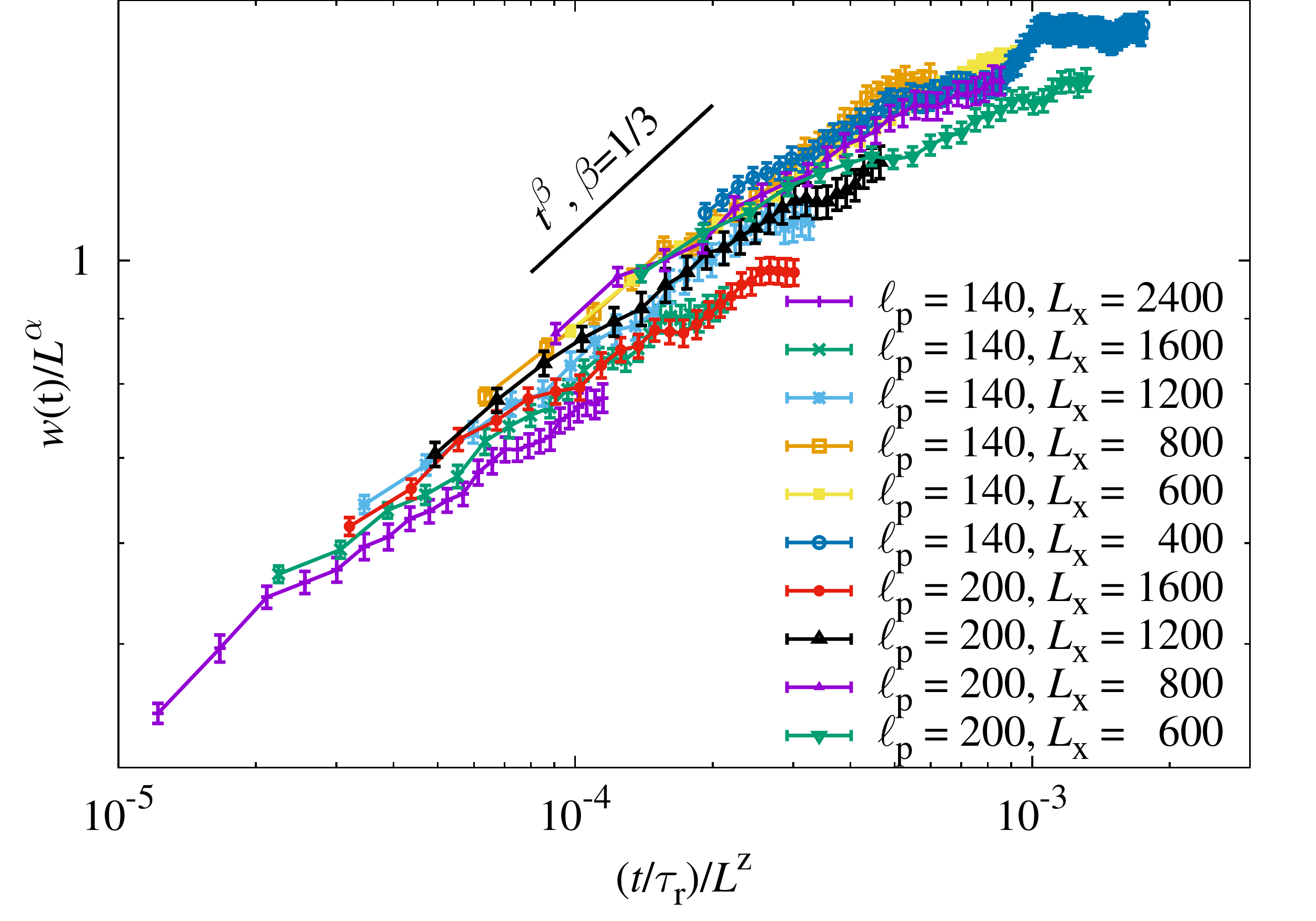} 
\caption{
The growth of the interfacial width $w(t)$   starting from a flat configuration
for several $\ell_\text{p}$ and system sizes is shown. We present a scaling
collapse according to the Edwards-Wilkinson (top) and Kardar-Parisi-Zhang
(bottom) critical exponents. Here $\alpha=1/2$, $z= \alpha/\beta$,
$\beta_\text{KPZ}=1/3$, $\beta_\text{EW}=1/4$.
\label{fig:scalingCollapse} }
\end{figure}

The key features of the MIPS interface seem to be a negative value of surface
tension coupled with a scale-free, nearly equilibrium-like spectrum of
interfacial fluctuations. The negative value of $\gamma$ suggests that either
additional terms in $\partial_t h$ or the coupling of $h$ to another field is
required. Natural candidates, such as $\nabla^4h$ terms or the coupling
of the interfacial dynamics to a scalar field describing the flux of particles
in and out of the two phases, select either a length scale or a time scale, and
neither is seen in our system (we have confirmed that in our data the power
spectrum of the interface $S(q,\omega)$ does not have any apparent time scale).
The slight deviations from the equilibrium scaling of $\langle |\delta
h(q)|^2\rangle$ may indicate that either a fundamentally nonlinear
phenomenological model is required, or that simulations of much larger systems
would reveal a long characteristic length scale in the problem. We view such
simulations as a natural object of future study to resolve this issue.

\section*{Appendices}
\appendix

\section{Locating the interface\label{App:contour}}
We have utilized two techniques to locate the interface and quantify its
properties. The first uses a global frame of reference with axes normal and
tangential to the mean (temporally and spatially averaged) interface ($x$ and
$y$) and yields a height map projected onto the $y$-axis, $h(y)$. The second
traces an outline of the boundary of the strip yielding a parameterized curve
$h(s)$ that captures fluctuations and overhangs. In both cases, for a given
snapshot we quantify the configuration and position of left and right
interface, denoted by $h_\text{L,R}$.

\subsection{Height map \boldmath $h(y)$}
To calculate the height map we follow a  straightforward procedure:
\begin{enumerate}
\item Distribute the cluster particles in $n_\text{bin}$ bins according to
their $y$-position.  
\item Sort particles in each bin according to their $x$-distance from the cluster center. 
\item Average this $x$-distance of the left-most or right-most.
$n_\text{avg}$ particles. The resulting quantity is identified with
$h_\text{L/R}(y)$. In general we use $n_\text{avg}=3$. 
\end{enumerate}
The width $d_y=L_y/n_\text{bins}$ of the bins is chosen as $d_y=2r_0$. We have
also verified that slightly larger bins do not significantly change our
measurements. 

\begin{figure}[t]
\includegraphics[height=0.45\textwidth,angle=270]{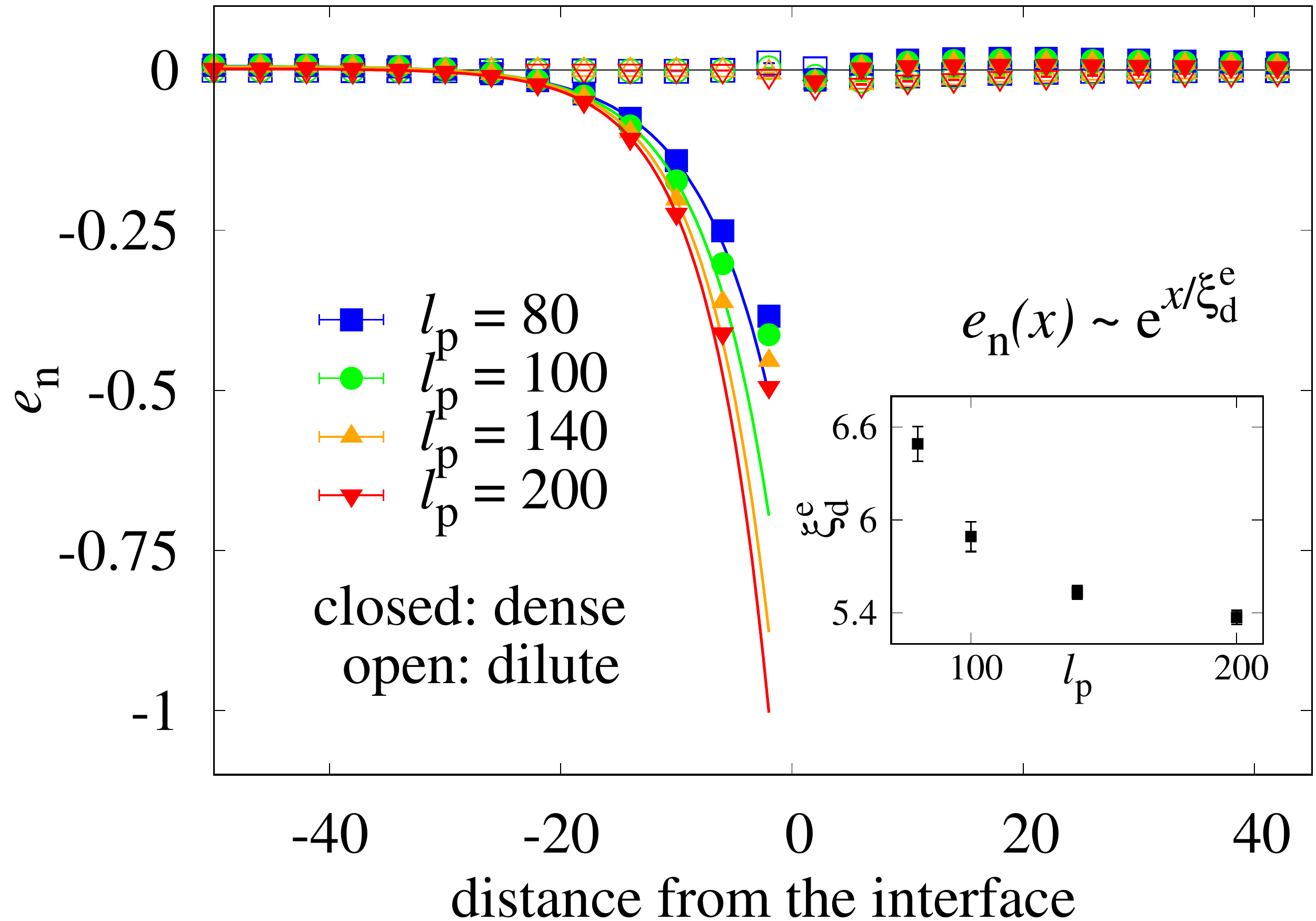} 
\caption{
The main figure shows the normal component of the polarization $e_\text{n}$
measured in the local frame and averaged along the interface. This is
proportional to the average local swim force per particle.  The solid (open)
symbols correspond to particles in the dense (gas) phase. The data show that
there is an excess of inwardly polarized particles in a surface layer of
thickness $\xi_\text{d}^\text{e}$ in the dense phase, but not in the gas.  The
length $\xi_\text{d}^\text{e}$ has been extracted with an exponential fit
to the decay of $e_\text{n}$ and is shown in the inset as a function of
$\ell_p$.  It is found to decay with increasing persistence.  The net
tangential component of polarization (not shown) remains zero at the
interface because particles travel without preference in either
direction tangent to the interface.
} 
\label{fig:en} 
\end{figure}

\subsection{Interface contour \boldmath $\tilde{h}(s)$}
We have also quantified the interface not as a height map but as a curve
parameterized by its arc length. To do so, we have implemented a classic
contour-finding algorithm\cite{toussaint} with square pixel resolution $l$
at the size of particles, $l=2r_0$. Schematically, the algorithm proceeds as
follows: 
\begin{enumerate}
\item Discretize the system into square pixels of size $l$. 
\item Mark all pixels containing cluster particles, call this set $G$.
\item Dilate $G$ by creating a new set $D$ such that $G \subset D$ where all pixels
adjacent to members of $G$ are marked in $D$.  
\item Determine set of contour pixels $C = D - G$. 
\item Connect contours using a depth-first-search algorithm, collating adjacent
points into subsets $c$ such that $c \in C$.
\end{enumerate}
The result can be seen in Fig.~\ref{fig:snapshot}. Once the sets $c$ are
available, we locate the longest contours that cross the periodic boundaries an
odd number of times. These are then sorted in  left and right according to
their $x$-distance from the strip center of mass $x_\text{CM}$. The height
$\tilde{h}(s)$ is then defined as the distance of these points from
$x_\text{CM}$.

\subsection{Smoothing}
While the above measure of the interface is generally useful for determining
the interface length, including the more fractal inlets that exist down to the
single-particle level, the noise on our measurements of local curvature is
significantly reduced when we smooth the interface contours $\tilde{h}(s)$ to
$h(s)$. To do this, we calculate the shortest path along $h(s)$ using the
Dijkstra algorithm\cite{cormen2009} on this relatively sparse, but connected set.
More specifically we 
\begin{enumerate}
\item Choose any pixel $p \in c$ with only two neighbor pixels $p_-,p_+ \in c$.
\item Run Dijkstra on reduced contour $(c - p)$, starting at $p_+$ and ending
at $p_-$. 
\item Add $p$ back to the Dijkstra path to connect the set $h(s)$. 
\end{enumerate} 
The resulting $h(s)$ is now a smooth connected path that winds the periodic box
in the $y$-direction.

\subsection{Determination of local normal to the interface}
Using either the basic contour $\tilde{h}(s)$ or the smoothed $h(s)$, we
calculate a local normal $\hat{\mathbf{n}}$ using the local tangent vectors, as
defined by each pixel and two nearest neighbors. We coarse-grain these vectors
by averaging the normal vectors of the $n_\text{cg}$ nearest neighbors contour
pixels. We test the result by eye and find that a coarse-graining of $n \approx
10$ neighboring pixels works well at several $\lp$. The result is a set of
normal vectors  $\mathbf{n}(s)$ along the interface. For consistency, we use
the same number of normals for the unsmoothed and smoothed interfaces, although
this number could be reduced in the smoothed case.

\begin{figure}[t]
\includegraphics[height=0.45\textwidth,angle=270]{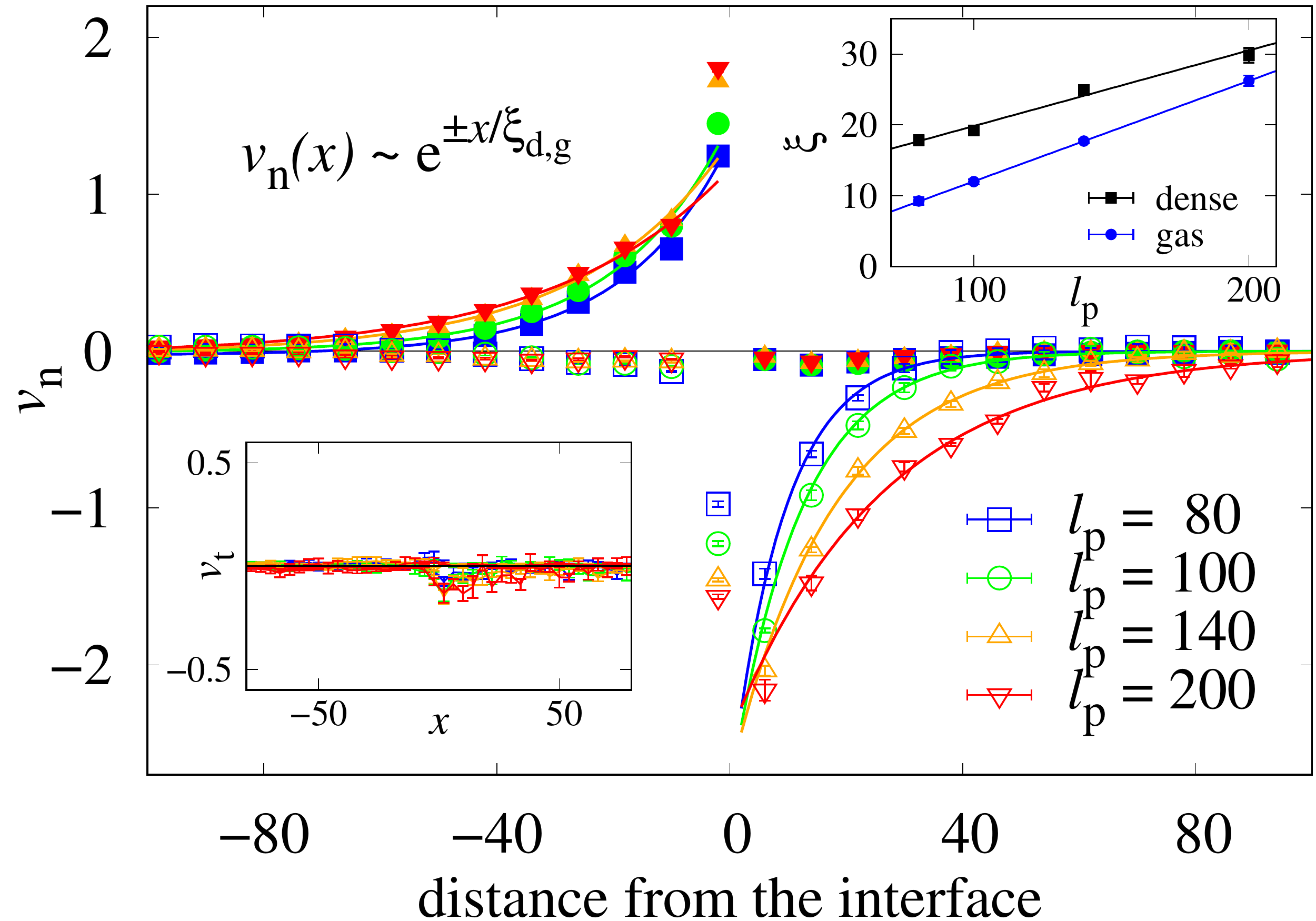} 
\caption{
The main figure shows the normal velocity $v_\text{n}$ of particles measured in
the local frame and averaged along the interface.  The solid (open) symbols
correspond to particles in the dense (gas) phase. The positive, outward-moving
average velocity just inside the dense phase is balanced by the negative
inward-moving average velocity just inside the gas phase. Within each phase,
the normal average velocity is finite within a surface layer of thickness
$\xi_\text{d}$ (dense) and $\xi_\text{g}$ (gas) and decays exponentially. The
solid lines are fits to the exponential decay.  The top right inset shows
$\xi_\text{d,g}$ extracted from those fits as functions of $\ell_p$.  The solid
lines are fits that show the linear growth of $\xi_\text{d,g}$ with
persistence. The bottom left inset shows that the signed  tangential component
of velocity is zero at the interface because particles travel without
preference in either direction tangent to the interface.  
}
\label{fig:vn} 
\end{figure}

\section{Local frame\label{App:localFrame}}
Here we provide additional details of how we defined the local frame of the
MIPS interface.

\subsection{Definition of frame}
The local frame is defined using $h(s)$ or $\tilde{h}(s)$ and the set of normal
vectors $\mathbf{n}(s)$, where $s$ is the set of connected points defining the
contour.  Contours $\tilde{h}(s)$ are multiple-valued functions, so when using
this contour we ignore those points that don't have enough neighbors to
properly define a normal. In the case of $\tilde{h}(s)$, this problem is
avoided by definition. 

Using $h(s)$ and $\textbf{n}(s)$, we define a set of bins contained in a box
of width $w_\text{c}$ length $L_\text{c}$, oriented along $\textbf{n}(s)$ and
centered at $h(s)$. The bin width $dn_\text{c}= 4 r_0$ sets the spatial
resolution of our local frame data. A schematic of this frame is shown in
Fig.~\ref{fig:snapshot}. With this binning technique, we collect information
for all particles that fall inside the box.  Given the local normal, it is
straightforward to properly transform the various scalar, vector and tensor
physical quantities measured in this work from the global ($x-y$) frame to the
local frame.

\section{Local dynamics near the interface}
\label{App:local_dynamics}

\balance

Previous work has focused on measuring the direction of the particles'
propulsive force (polarization), identifying the excess of particles with
inward polarization just inside the dense cluster as the main stabilizing
mechanism. In this paper, we have examined in more detail the local dynamics in
the interfacial region both within the dense and the gas phases. While we do
observe an excess of particles with inward pointing polarization at the surface
of the dense phase (see Fig.~\ref{fig:en}), the most intriguing observation is
of sustained local \textit{tangential} motion of particles in \textit{both}
regions. This dynamics is displayed in more detail in Fig.\ref{fig:vn}, where
we show that the anisotropy of motion extends into each phase within a surface
layer. Additionally, while the thickness of the region of inward-pointing
\textit{polarization} in the dense phase decreases with increasing persistence
(see inset of Fig.~\ref{fig:en}), the thickness of the layers with finite
particles \textit{velocity}  increases with persistence in both the gas and the
dense phases.  In these regions we observe both normal currents that balance
each other and tangential currents that average to zero (Fig.~\ref{fig:vn}),
but greatly exceed in magnitude the normal currents
(Fig.~\ref{fig:local_dynamics}).  This suggests that the stiffening of the
interface results from the combined effect of inward polarization excess at the
cluster's surface,  local tangential flows in the gas, and the self-shearing of
the interface in the dense cluster that heals fluctuations and enhances
stability. This observation may provide additional intuition in constructing
simple models of MIPS systems, such as the Active Matter Model
B+\cite{tjhung2018reverse}.

\begin{figure}[t]
\includegraphics[height=0.45\textwidth,angle=270]{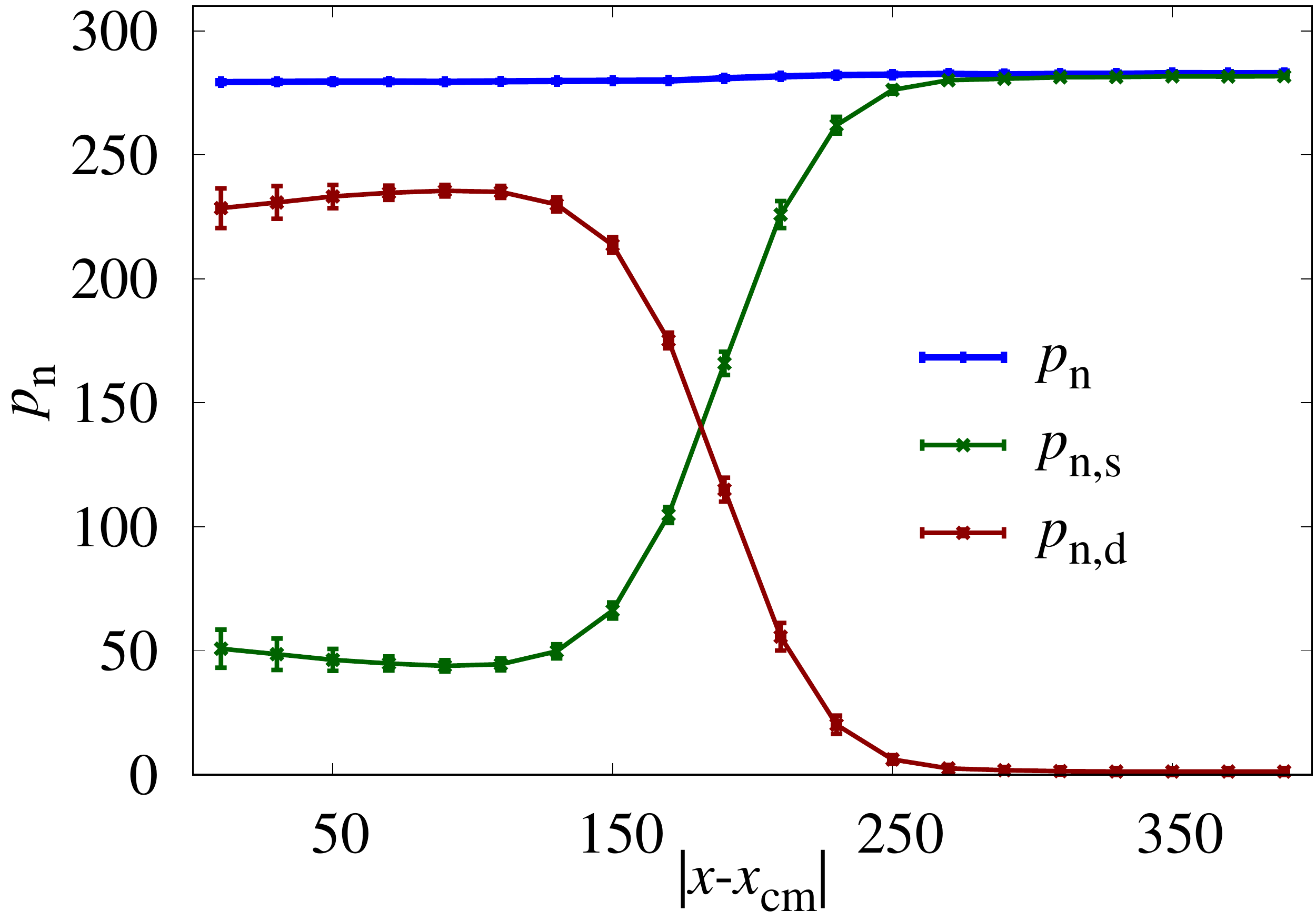} 
\includegraphics[height=0.45\textwidth,angle=270]{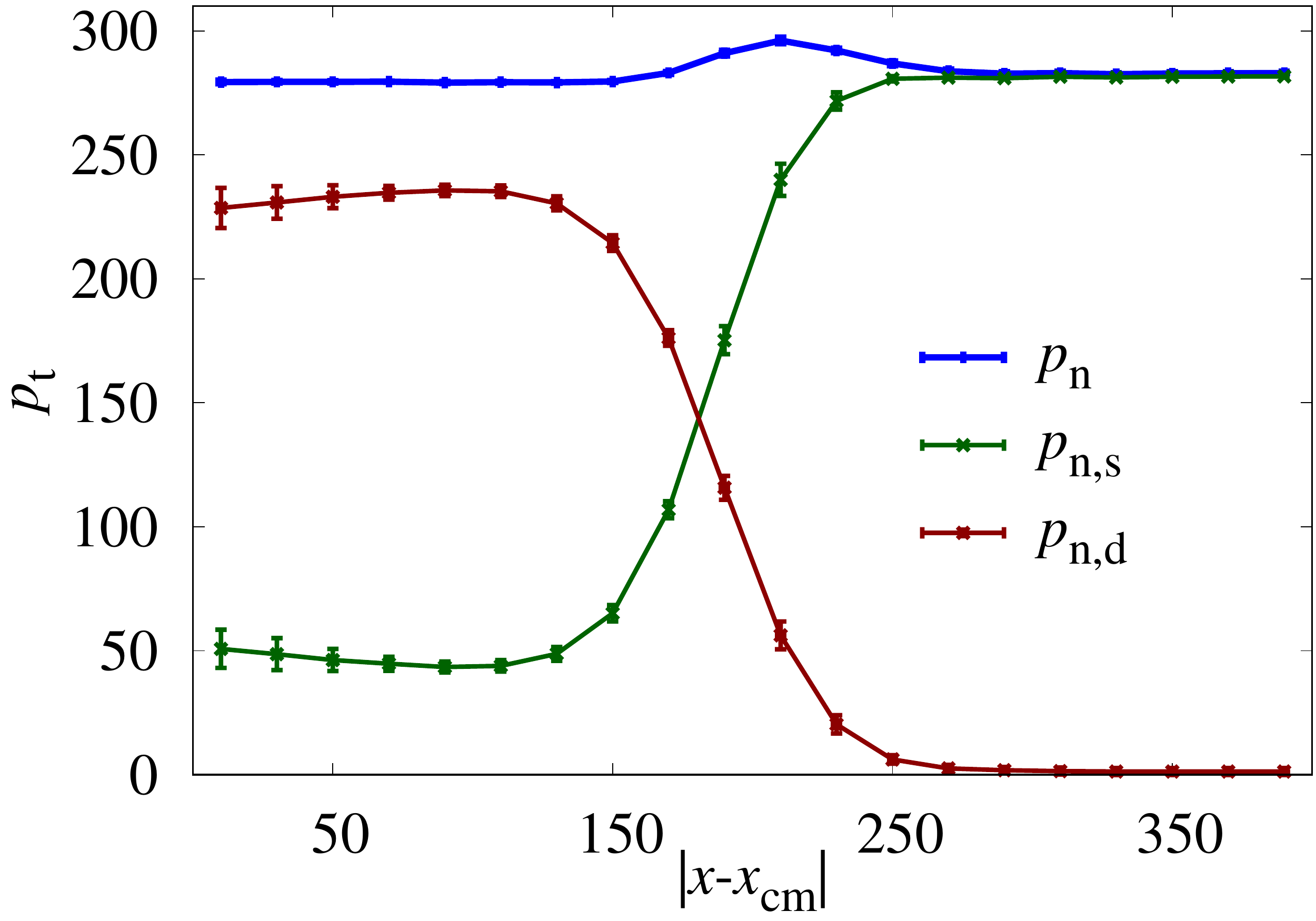} 
\includegraphics[height=0.45\textwidth,angle=270]{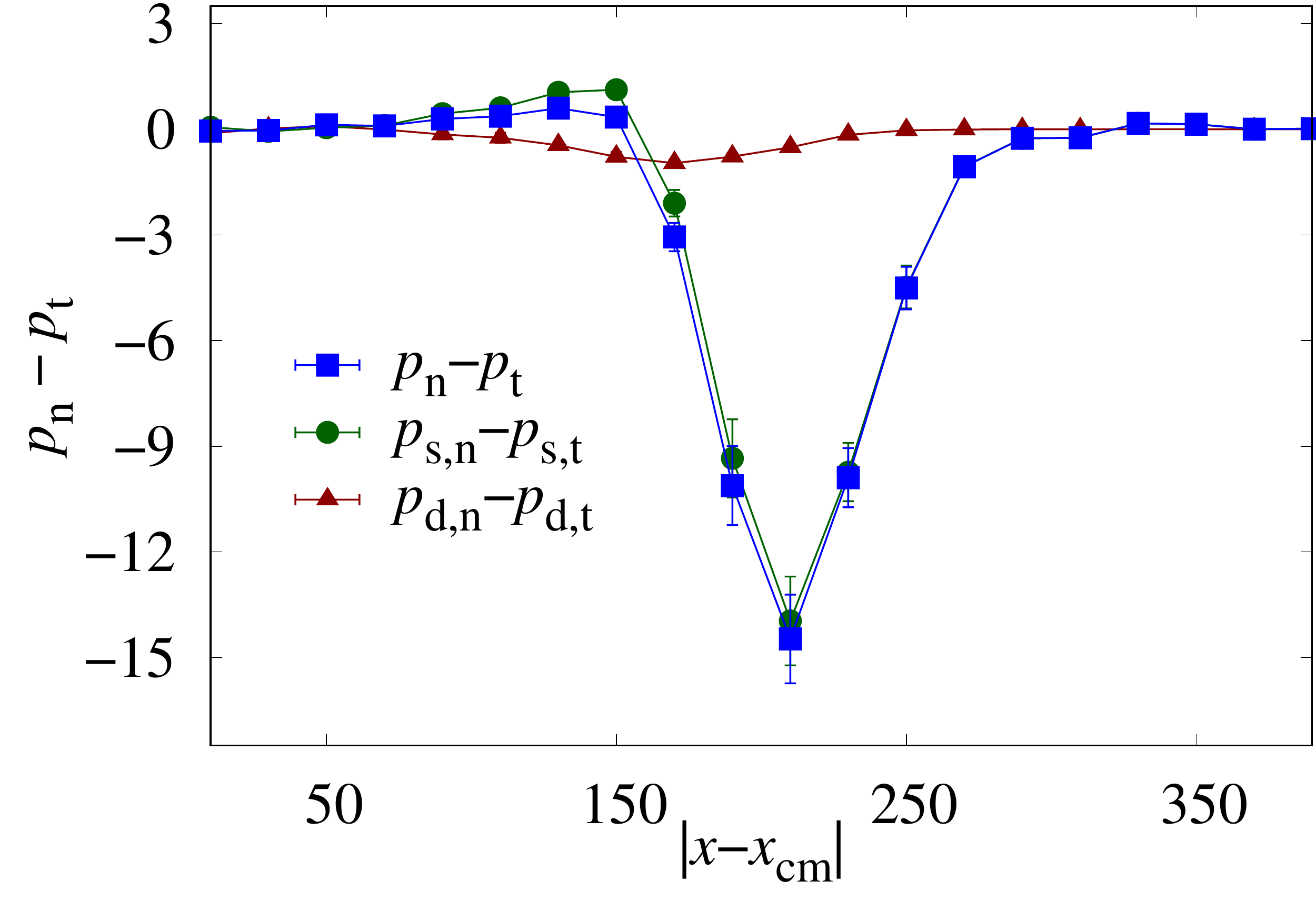} 
\caption{
Normal (top) and tangential (middle) pressure broken down into swim,
interaction and total parts. The bottom panel shows the various
contributions (total, swim and interaction) to the difference $p_\text{n} -
p_\text{t}$ as a function of the distance $x$ from the dense phase center of
mass, for $\ell_\text{p}=140$. This difference is the integrand in the
expression for the tension, Eq.~\eqref{eq:tension}.  
\label{fig:breakdown}
}
\end{figure}

\section*{Conflicts of interest}
%In accordance with our policy on \href{http://www.rsc.org/journals-books-databases/journal-authors-reviewers/author-responsibilities/#code-of-conduct}{Conflicts of interest} 
There are no conflicts to declare.

\section*{Acknowledgments}
We thank Suraj Shankar and Michael Cates for helpful discussions. This work was
primarily supported by NSF-DMR-1609208. Additional support was provided by
NSF-POLS-1607416 (DMS), NSF-PHY-1748958 (DY, MCM) and  FIS2015-65078-C2-1-P
(DY), jointly funded by MINECO (Spain) and FEDER (EU). MCM and AP acknowledge
support by the NSF IGERT program through award NSF-DGE-1068780.  All authors
acknowledge support of the Syracuse University Soft and Living Matter Program.
Our simulations were carried out on a Tesla K40 donated by the NVIDIA
corporation, on the Cierzo supercomputer (BIFI-ZCAM, Universidad de Zaragoza)
and on the Syracuse University HTC Campus Grid, which is supported by
NSF-ACI-1341006.  MCM and DY thank the KITP for hospitality during part of this
project.

%%%END OF MAIN TEXT%%%


\balance


%%%REFERENCES%%%
%\bibliography{mips-interface} %You need to replace "rsc" on this line with the name of your .bib file

\begin{mcitethebibliography}{43}
\providecommand*{\natexlab}[1]{#1}
\providecommand*{\mciteSetBstSublistMode}[1]{}
\providecommand*{\mciteSetBstMaxWidthForm}[2]{}
\providecommand*{\mciteBstWouldAddEndPuncttrue}
  {\def\EndOfBibitem{\unskip.}}
\providecommand*{\mciteBstWouldAddEndPunctfalse}
  {\let\EndOfBibitem\relax}
\providecommand*{\mciteSetBstMidEndSepPunct}[3]{}
\providecommand*{\mciteSetBstSublistLabelBeginEnd}[3]{}
\providecommand*{\EndOfBibitem}{}
\mciteSetBstSublistMode{f}
\mciteSetBstMaxWidthForm{subitem}
{(\emph{\alph{mcitesubitemcount}})}
\mciteSetBstSublistLabelBeginEnd{\mcitemaxwidthsubitemform\space}
{\relax}{\relax}

\bibitem[de~Gennes(2005)]{de2005soft}
P.~G. de~Gennes, \emph{Soft interfaces: the 1994 Dirac memorial lecture},
  Cambridge University Press, 2005\relax
\mciteBstWouldAddEndPuncttrue
\mciteSetBstMidEndSepPunct{\mcitedefaultmidpunct}
{\mcitedefaultendpunct}{\mcitedefaultseppunct}\relax
\EndOfBibitem
\bibitem[Fily and Marchetti(2012)]{Fily2012a}
Y.~Fily and M.~C. Marchetti, \emph{Physical Review Letters}, 2012,
  \textbf{108}, 1--5\relax
\mciteBstWouldAddEndPuncttrue
\mciteSetBstMidEndSepPunct{\mcitedefaultmidpunct}
{\mcitedefaultendpunct}{\mcitedefaultseppunct}\relax
\EndOfBibitem
\bibitem[Tailleur and Cates(2008)]{Tailleur2008}
J.~Tailleur and M.~E. Cates, \emph{Phys. Rev. Lett.}, 2008, \textbf{100},
  218103\relax
\mciteBstWouldAddEndPuncttrue
\mciteSetBstMidEndSepPunct{\mcitedefaultmidpunct}
{\mcitedefaultendpunct}{\mcitedefaultseppunct}\relax
\EndOfBibitem
\bibitem[Redner \emph{et~al.}(2013)Redner, Baskaran, and Hagan]{Redner2013}
G.~S. Redner, A.~Baskaran and M.~F. Hagan, \emph{Physical Review E}, 2013,
  \textbf{88}, 012305\relax
\mciteBstWouldAddEndPuncttrue
\mciteSetBstMidEndSepPunct{\mcitedefaultmidpunct}
{\mcitedefaultendpunct}{\mcitedefaultseppunct}\relax
\EndOfBibitem
\bibitem[Cates and Tailleur(2015)]{Cates2015}
M.~E. Cates and J.~Tailleur, \emph{Annual Review of Condensed Matter Physics},
  2015, \textbf{6}, 150112144536003\relax
\mciteBstWouldAddEndPuncttrue
\mciteSetBstMidEndSepPunct{\mcitedefaultmidpunct}
{\mcitedefaultendpunct}{\mcitedefaultseppunct}\relax
\EndOfBibitem
\bibitem[Wittkowski \emph{et~al.}(2014)Wittkowski, Tiribocchi, Stenhammar,
  Allen, Marenduzzo, and Cates]{Wittkowski2014}
R.~Wittkowski, A.~Tiribocchi, J.~Stenhammar, R.~J. Allen, D.~Marenduzzo and
  M.~E. Cates, \emph{Nature communications}, 2014, \textbf{5}, 4351\relax
\mciteBstWouldAddEndPuncttrue
\mciteSetBstMidEndSepPunct{\mcitedefaultmidpunct}
{\mcitedefaultendpunct}{\mcitedefaultseppunct}\relax
\EndOfBibitem
\bibitem[Fily \emph{et~al.}(2014)Fily, Henkes, and Marchetti]{Fily2014}
Y.~Fily, S.~Henkes and M.~C. Marchetti, \emph{Soft matter}, 2014, \textbf{10},
  2132--40\relax
\mciteBstWouldAddEndPuncttrue
\mciteSetBstMidEndSepPunct{\mcitedefaultmidpunct}
{\mcitedefaultendpunct}{\mcitedefaultseppunct}\relax
\EndOfBibitem
\bibitem[Solon \emph{et~al.}(2015)Solon, Stenhammar, Wittkowski, Kardar, Kafri,
  Cates, and Tailleur]{Solon2015}
A.~P. Solon, J.~Stenhammar, R.~Wittkowski, M.~Kardar, Y.~Kafri, M.~E. Cates and
  J.~Tailleur, \emph{Physical Review Letters}, 2015, \textbf{198301},
  1--6\relax
\mciteBstWouldAddEndPuncttrue
\mciteSetBstMidEndSepPunct{\mcitedefaultmidpunct}
{\mcitedefaultendpunct}{\mcitedefaultseppunct}\relax
\EndOfBibitem
\bibitem[Solon \emph{et~al.}(2015)Solon, Fily, Baskaran, Cates, Kafri, Kardar,
  and Tailleur]{Solon2015a}
A.~P. Solon, Y.~Fily, A.~Baskaran, M.~E. Cates, Y.~Kafri, M.~Kardar and
  J.~Tailleur, \emph{Nature Physics}, 2015, \textbf{11}, 673--678\relax
\mciteBstWouldAddEndPuncttrue
\mciteSetBstMidEndSepPunct{\mcitedefaultmidpunct}
{\mcitedefaultendpunct}{\mcitedefaultseppunct}\relax
\EndOfBibitem
\bibitem[Winkler \emph{et~al.}(2015)Winkler, Wysocki, and Gompper]{Winkler2015}
R.~G. Winkler, A.~Wysocki and G.~Gompper, \emph{Soft Matter}, 2015,
  \textbf{11}, 6680--6691\relax
\mciteBstWouldAddEndPuncttrue
\mciteSetBstMidEndSepPunct{\mcitedefaultmidpunct}
{\mcitedefaultendpunct}{\mcitedefaultseppunct}\relax
\EndOfBibitem
\bibitem[Takatori and Brady(2015)]{Takatori2015}
S.~C. Takatori and J.~F. Brady, \emph{Soft Matter}, 2015\relax
\mciteBstWouldAddEndPuncttrue
\mciteSetBstMidEndSepPunct{\mcitedefaultmidpunct}
{\mcitedefaultendpunct}{\mcitedefaultseppunct}\relax
\EndOfBibitem
\bibitem[{Marini Bettolo Marconi} \emph{et~al.}(2016){Marini Bettolo Marconi},
  Maggi, and Melchionna]{marconi2016}
U.~{Marini Bettolo Marconi}, C.~Maggi and S.~Melchionna, \emph{Soft Matter},
  2016, \textbf{12}, 5727--5738\relax
\mciteBstWouldAddEndPuncttrue
\mciteSetBstMidEndSepPunct{\mcitedefaultmidpunct}
{\mcitedefaultendpunct}{\mcitedefaultseppunct}\relax
\EndOfBibitem
\bibitem[Solon \emph{et~al.}(2018)Solon, Stenhammar, Cates, Kafri, and
  Tailleur]{solon2018generalized}
A.~P. Solon, J.~Stenhammar, M.~E. Cates, Y.~Kafri and J.~Tailleur, \emph{arXiv
  preprint arXiv:1803.06159}, 2018\relax
\mciteBstWouldAddEndPuncttrue
\mciteSetBstMidEndSepPunct{\mcitedefaultmidpunct}
{\mcitedefaultendpunct}{\mcitedefaultseppunct}\relax
\EndOfBibitem
\bibitem[Redner \emph{et~al.}(2013)Redner, Hagan, and Baskaran]{Redner2013b}
G.~S. Redner, M.~F. Hagan and A.~Baskaran, \emph{Physical Review Letters},
  2013, \textbf{110}, 1--5\relax
\mciteBstWouldAddEndPuncttrue
\mciteSetBstMidEndSepPunct{\mcitedefaultmidpunct}
{\mcitedefaultendpunct}{\mcitedefaultseppunct}\relax
\EndOfBibitem
\bibitem[Bialk{\'{e}} \emph{et~al.}(2013)Bialk{\'{e}}, L{\"{o}}wen, and
  Speck]{Bialke2013}
J.~Bialk{\'{e}}, H.~L{\"{o}}wen and T.~Speck, \emph{EPL (Europhysics Letters)},
  2013, \textbf{103}, 30008\relax
\mciteBstWouldAddEndPuncttrue
\mciteSetBstMidEndSepPunct{\mcitedefaultmidpunct}
{\mcitedefaultendpunct}{\mcitedefaultseppunct}\relax
\EndOfBibitem
\bibitem[Yang \emph{et~al.}(2014)Yang, Manning, and Marchetti]{Yang2014}
X.~Yang, M.~L. Manning and M.~C. Marchetti, \emph{Soft matter}, 2014,
  \textbf{672}, 6477--6484\relax
\mciteBstWouldAddEndPuncttrue
\mciteSetBstMidEndSepPunct{\mcitedefaultmidpunct}
{\mcitedefaultendpunct}{\mcitedefaultseppunct}\relax
\EndOfBibitem
\bibitem[Takatori and Brady(2015)]{takatori2015thermo}
S.~C. Takatori and J.~F. Brady, \emph{Phys. Rev. E}, 2015, \textbf{91},
  032117\relax
\mciteBstWouldAddEndPuncttrue
\mciteSetBstMidEndSepPunct{\mcitedefaultmidpunct}
{\mcitedefaultendpunct}{\mcitedefaultseppunct}\relax
\EndOfBibitem
\bibitem[Theurkauff \emph{et~al.}(2012)Theurkauff, Cottin-Bizonne, Palacci,
  Ybert, and Bocquet]{theurkauff2012}
I.~Theurkauff, C.~Cottin-Bizonne, J.~Palacci, C.~Ybert and L.~Bocquet,
  \emph{Phys. Rev. Lett.}, 2012, \textbf{108}, 268303\relax
\mciteBstWouldAddEndPuncttrue
\mciteSetBstMidEndSepPunct{\mcitedefaultmidpunct}
{\mcitedefaultendpunct}{\mcitedefaultseppunct}\relax
\EndOfBibitem
\bibitem[Buttinoni \emph{et~al.}(2013)Buttinoni, Bialk\'e, K\"ummel, L\"owen,
  Bechinger, and Speck]{Buttinoni2013}
I.~Buttinoni, J.~Bialk\'e, F.~K\"ummel, H.~L\"owen, C.~Bechinger and T.~Speck,
  \emph{Phys. Rev. Lett.}, 2013, \textbf{110}, 238301\relax
\mciteBstWouldAddEndPuncttrue
\mciteSetBstMidEndSepPunct{\mcitedefaultmidpunct}
{\mcitedefaultendpunct}{\mcitedefaultseppunct}\relax
\EndOfBibitem
\bibitem[Palacci \emph{et~al.}(2013)Palacci, Sacanna, Steinberg, Pine, and
  Chaikin]{Palacci2013}
J.~Palacci, S.~Sacanna, A.~P. Steinberg, D.~J. Pine and P.~M. Chaikin,
  \emph{Science}, 2013, \textbf{339}, 936--940\relax
\mciteBstWouldAddEndPuncttrue
\mciteSetBstMidEndSepPunct{\mcitedefaultmidpunct}
{\mcitedefaultendpunct}{\mcitedefaultseppunct}\relax
\EndOfBibitem
\bibitem[Mallory \emph{et~al.}(2014)Mallory, \ifmmode \check{S}\else
  \v{S}\fi{}ari\ifmmode~\acute{c}\else \'{c}\fi{}, Valeriani, and
  Cacciuto]{mallory2014}
S.~A. Mallory, A.~\ifmmode \check{S}\else \v{S}\fi{}ari\ifmmode~\acute{c}\else
  \'{c}\fi{}, C.~Valeriani and A.~Cacciuto, \emph{Phys. Rev. E}, 2014,
  \textbf{89}, 052303\relax
\mciteBstWouldAddEndPuncttrue
\mciteSetBstMidEndSepPunct{\mcitedefaultmidpunct}
{\mcitedefaultendpunct}{\mcitedefaultseppunct}\relax
\EndOfBibitem
\bibitem[Bechinger \emph{et~al.}(2016)Bechinger, Di~Leonardo, L\"owen,
  Reichhardt, Volpe, and Volpe]{Bechinger2016}
C.~Bechinger, R.~Di~Leonardo, H.~L\"owen, C.~Reichhardt, G.~Volpe and G.~Volpe,
  \emph{Rev. Mod. Phys.}, 2016, \textbf{88}, 045006\relax
\mciteBstWouldAddEndPuncttrue
\mciteSetBstMidEndSepPunct{\mcitedefaultmidpunct}
{\mcitedefaultendpunct}{\mcitedefaultseppunct}\relax
\EndOfBibitem
\bibitem[Liu \emph{et~al.}(2017)Liu, Patch, Bahar, Yllanes, Welch, Marchetti,
  Thutupalli, and Shaevitz]{Liu2017}
G.~Liu, A.~Patch, F.~Bahar, D.~Yllanes, R.~D. Welch, M.~C. Marchetti,
  S.~Thutupalli and J.~W. Shaevitz, \emph{arXiv:1709.06012}, 2017\relax
\mciteBstWouldAddEndPuncttrue
\mciteSetBstMidEndSepPunct{\mcitedefaultmidpunct}
{\mcitedefaultendpunct}{\mcitedefaultseppunct}\relax
\EndOfBibitem
\bibitem[Bialk{\'{e}} \emph{et~al.}(2015)Bialk{\'{e}}, Siebert, L{\"{o}}wen,
  and Speck]{Bialke2015}
J.~Bialk{\'{e}}, J.~T. Siebert, H.~L{\"{o}}wen and T.~Speck, \emph{Physical
  Review E}, 2015, \textbf{098301}, 1--5\relax
\mciteBstWouldAddEndPuncttrue
\mciteSetBstMidEndSepPunct{\mcitedefaultmidpunct}
{\mcitedefaultendpunct}{\mcitedefaultseppunct}\relax
\EndOfBibitem
\bibitem[Speck(2016)]{speck2016stochastic}
T.~Speck, \emph{EPL (Europhysics Letters)}, 2016, \textbf{114}, 30006\relax
\mciteBstWouldAddEndPuncttrue
\mciteSetBstMidEndSepPunct{\mcitedefaultmidpunct}
{\mcitedefaultendpunct}{\mcitedefaultseppunct}\relax
\EndOfBibitem
\bibitem[Lee(2017)]{lee2017interface}
C.~F. Lee, \emph{Soft matter}, 2017, \textbf{13}, 376--385\relax
\mciteBstWouldAddEndPuncttrue
\mciteSetBstMidEndSepPunct{\mcitedefaultmidpunct}
{\mcitedefaultendpunct}{\mcitedefaultseppunct}\relax
\EndOfBibitem
\bibitem[Paliwal \emph{et~al.}(2017)Paliwal, Prymidis, Filion, and
  Dijkstra]{paliwal2017}
S.~Paliwal, V.~Prymidis, L.~Filion and M.~Dijkstra, \emph{The Journal of
  Chemical Physics}, 2017, \textbf{147}, 84902\relax
\mciteBstWouldAddEndPuncttrue
\mciteSetBstMidEndSepPunct{\mcitedefaultmidpunct}
{\mcitedefaultendpunct}{\mcitedefaultseppunct}\relax
\EndOfBibitem
\bibitem[Tjhung \emph{et~al.}(2018)Tjhung, Nardini, and
  Cates]{tjhung2018reverse}
E.~Tjhung, C.~Nardini and M.~E. Cates, \emph{arXiv preprint arXiv:1801.07687},
  2018\relax
\mciteBstWouldAddEndPuncttrue
\mciteSetBstMidEndSepPunct{\mcitedefaultmidpunct}
{\mcitedefaultendpunct}{\mcitedefaultseppunct}\relax
\EndOfBibitem
\bibitem[Edwards \emph{et~al.}(1982)Edwards,
  Wilkinson,\emph{et~al.}]{edwards1982surface}
S.~F. Edwards, D.~Wilkinson \emph{et~al.}, Proc. R. Soc. Lond. A, 1982, pp.
  17--31\relax
\mciteBstWouldAddEndPuncttrue
\mciteSetBstMidEndSepPunct{\mcitedefaultmidpunct}
{\mcitedefaultendpunct}{\mcitedefaultseppunct}\relax
\EndOfBibitem
\bibitem[Kardar \emph{et~al.}(1986)Kardar, Parisi, and Zhang]{KPZ}
M.~Kardar, G.~Parisi and Y.-C. Zhang, \emph{Physical Review Letters}, 1986,
  \textbf{56}, 889\relax
\mciteBstWouldAddEndPuncttrue
\mciteSetBstMidEndSepPunct{\mcitedefaultmidpunct}
{\mcitedefaultendpunct}{\mcitedefaultseppunct}\relax
\EndOfBibitem
\bibitem[Marchetti \emph{et~al.}(2016)Marchetti, Fily, Henkes, Patch, and
  Yllanes]{marchetti2016minimal}
M.~C. Marchetti, Y.~Fily, S.~Henkes, A.~Patch and D.~Yllanes, \emph{Current
  Opinion in Colloid \& Interface Science}, 2016, \textbf{21}, 34--43\relax
\mciteBstWouldAddEndPuncttrue
\mciteSetBstMidEndSepPunct{\mcitedefaultmidpunct}
{\mcitedefaultendpunct}{\mcitedefaultseppunct}\relax
\EndOfBibitem
\bibitem[Rowlinson and Widom(2013)]{rowlinson2013molecular}
J.~S. Rowlinson and B.~Widom, \emph{Molecular theory of capillarity}, Courier
  Corporation, 2013\relax
\mciteBstWouldAddEndPuncttrue
\mciteSetBstMidEndSepPunct{\mcitedefaultmidpunct}
{\mcitedefaultendpunct}{\mcitedefaultseppunct}\relax
\EndOfBibitem
\bibitem[Kirkwood and Buff(1949)]{kirkwood1949statistical}
J.~G. Kirkwood and F.~P. Buff, \emph{The Journal of Chemical Physics}, 1949,
  \textbf{17}, 338--343\relax
\mciteBstWouldAddEndPuncttrue
\mciteSetBstMidEndSepPunct{\mcitedefaultmidpunct}
{\mcitedefaultendpunct}{\mcitedefaultseppunct}\relax
\EndOfBibitem
\bibitem[Patch \emph{et~al.}(2017)Patch, Yllanes, and
  Marchetti]{patch2017kinetics}
A.~Patch, D.~Yllanes and M.~C. Marchetti, \emph{Physical Review E}, 2017,
  \textbf{95}, 012601\relax
\mciteBstWouldAddEndPuncttrue
\mciteSetBstMidEndSepPunct{\mcitedefaultmidpunct}
{\mcitedefaultendpunct}{\mcitedefaultseppunct}\relax
\EndOfBibitem
\bibitem[Fily \emph{et~al.}(2017)Fily, Kafri, Solon, Tailleur, and
  Turner]{fily2017mechanical}
Y.~Fily, Y.~Kafri, A.~P. Solon, J.~Tailleur and A.~Turner, \emph{Journal of
  Physics A: Mathematical and Theoretical}, 2017, \textbf{51}, 044003\relax
\mciteBstWouldAddEndPuncttrue
\mciteSetBstMidEndSepPunct{\mcitedefaultmidpunct}
{\mcitedefaultendpunct}{\mcitedefaultseppunct}\relax
\EndOfBibitem
\bibitem[Takatori and Brady(2014)]{Takatori2014}
S.~C. Takatori and J.~F. Brady, \emph{Soft matter}, 2014,  9433--9445\relax
\mciteBstWouldAddEndPuncttrue
\mciteSetBstMidEndSepPunct{\mcitedefaultmidpunct}
{\mcitedefaultendpunct}{\mcitedefaultseppunct}\relax
\EndOfBibitem
\bibitem[Anderson \emph{et~al.}(2008)Anderson, Lorenz, and
  Travesset]{hoomd-blue}
J.~A. Anderson, C.~D. Lorenz and A.~Travesset, \emph{J. Comput. Phys.}, 2008,
  \textbf{227}, 5342\relax
\mciteBstWouldAddEndPuncttrue
\mciteSetBstMidEndSepPunct{\mcitedefaultmidpunct}
{\mcitedefaultendpunct}{\mcitedefaultseppunct}\relax
\EndOfBibitem
\bibitem[hoo(2017)]{hoomd-blue0}
\emph{HOOMD-blue}, \url{https://codeblue.umich.edu/hoomd-blue/}, 2017\relax
\mciteBstWouldAddEndPuncttrue
\mciteSetBstMidEndSepPunct{\mcitedefaultmidpunct}
{\mcitedefaultendpunct}{\mcitedefaultseppunct}\relax
\EndOfBibitem
\bibitem[del Junco and Vaikuntanathan(2018)]{Vaikuntanathan2018}
C.~del Junco and S.~Vaikuntanathan, \emph{arXiv:1803.02678}, 2018\relax
\mciteBstWouldAddEndPuncttrue
\mciteSetBstMidEndSepPunct{\mcitedefaultmidpunct}
{\mcitedefaultendpunct}{\mcitedefaultseppunct}\relax
\EndOfBibitem
\bibitem[Fily \emph{et~al.}(2014)Fily, Baskaran, and Hagan]{Fily2014a}
Y.~Fily, A.~Baskaran and M.~F. Hagan, \emph{Soft Matter}, 2014, \textbf{10},
  5609--5617\relax
\mciteBstWouldAddEndPuncttrue
\mciteSetBstMidEndSepPunct{\mcitedefaultmidpunct}
{\mcitedefaultendpunct}{\mcitedefaultseppunct}\relax
\EndOfBibitem
\bibitem[Fily \emph{et~al.}(2017)Fily, Baskaran, and
  Hagan]{fily2017equilibrium}
Y.~Fily, A.~Baskaran and M.~F. Hagan, \emph{The European Physical Journal E},
  2017, \textbf{40}, 61\relax
\mciteBstWouldAddEndPuncttrue
\mciteSetBstMidEndSepPunct{\mcitedefaultmidpunct}
{\mcitedefaultendpunct}{\mcitedefaultseppunct}\relax
\EndOfBibitem
\bibitem[Toussaint()]{toussaint}
G.~Toussaint, \emph{Grids, Connectivity and Contour Tracing}, accessed: 1
  November, 2017\relax
\mciteBstWouldAddEndPuncttrue
\mciteSetBstMidEndSepPunct{\mcitedefaultmidpunct}
{\mcitedefaultendpunct}{\mcitedefaultseppunct}\relax
\EndOfBibitem
\bibitem[Cormen \emph{et~al.}(2009)Cormen, Leiserson, Rivest, and
  Stein]{cormen2009}
T.~Cormen, C.~Leiserson, R.~Rivest and C.~Stein, \emph{Introduction to
  Algorithms}, MIT Press, 3rd edn, 2009\relax
\mciteBstWouldAddEndPuncttrue
\mciteSetBstMidEndSepPunct{\mcitedefaultmidpunct}
{\mcitedefaultendpunct}{\mcitedefaultseppunct}\relax
\EndOfBibitem
\end{mcitethebibliography}
\bibliographystyle{rsc} %the RSC's .bst file

\providecommand*{\mcitethebibliography}{\thebibliography}
\csname @ifundefined\endcsname{endmcitethebibliography}
{\let\endmcitethebibliography\endthebibliography}{}

\end{document}